\documentclass[aps,prb,twocolumn,superscriptaddress,showpacs,english]{revtex4-1}

\usepackage[T1]{fontenc}
\usepackage[latin9]{inputenc}
\usepackage{babel,amsmath,amssymb,wasysym,graphicx,xcolor}
\usepackage[linktocpage=true,
colorlinks=true, 
pdfborder={0 0 0},
linkcolor=blue,
citecolor=blue,
filecolor=yellow,
urlcolor=blue,
bookmarks,
pdfauthor={},
]{hyperref}

\bibpunct{[}{]}{,}{n}{}{}

\newcommand{\Oxford}{Department of Materials, University of Oxford, Parks Road, Oxford OX1 3PH, 
United Kingdom}
\newcommand{\TUGraz}{Institute of Theoretical and Computational Physics, Graz University of Technology, NAWI Graz, 8010 Graz, Austria}

\usepackage{soul}

\begin{document}

\title{Quasiparticle $GW$ band structures and Fermi surfaces of bulk and monolayer NbS$_2$}

\author{Christoph Heil}     \affiliation{\Oxford} \affiliation{\TUGraz}
\author{Martin Schlipf}   \affiliation{\Oxford}
\author{Feliciano Giustino} \email{feliciano.giustino@materials.ox.ac.uk} \affiliation{\Oxford}
\affiliation{Department of Materials Science and Engineering, Cornell University, Ithaca, New York 14853, USA}

\date{\today}

\begin{abstract}
In this work we employ the $GW$ approximation in the framework of the SternheimerGW method to investigate the effects of many-body corrections to the band structures and Fermi surfaces of bulk and monolayer NbS$_2$. For the bulk system, we find that the inclusion of these many-body effects leads to important changes in the band structure, especially in the low-energy regime around the Fermi level, and that our calculations are in good agreement with recent ARPES measurements. In the case of a free-standing monolayer NbS$_2$, we observe a strong increase of the screened Coulomb interaction and the quasiparticle corrections as compared to bulk. In this case we also perform calculations to include the effect of screening by a substrate. We report in detail the results of our convergence tests and computational parameters, to serve as a solid basis for future studies.
\end{abstract}

\maketitle

\section{Introduction}
Transition metal dichalcogenides (TMDs) have been the focus of many studies in recent years, as their physical and chemical diversity offers an ideal platform to investigate semiconductors, metals, and superconductors in layered systems using the same structural template~\cite{sipos_mott_2008,wang_electronics_2012,geim_van_2013,manzeli_2d_2017}. 
Within the family of TMDs, metallic materials have attracted considerable attention due to the fact that both a superconducting phase as well as a charge density wave (CDW) phase appear in the low-temperature phase diagram~\cite{wilson_transition_1969,moncton_study_1975,valla_quasiparticle_2004,weber_extended_2011,calandra_charge-density_2011,rosner_phase_2014,leroux_strong_2015,das_superconducting_2015,klemm_pristine_2015}. 
Moreover, the fact that the increased spatial confinement in few-layer and monolayer systems, as well as the substrates they are placed on, cause important changes in the electronic structure~\cite{DING20112254}, in particular in the screened Coulomb interaction, has drawn considerable interest to two-dimensional (2D) materials~\cite{park_first-principles_2009,das_sarma_electronic_2011,kotov_electron-electron_2012,qiu_optical_2013,ugeda_giant_2014,yu_gate-tunable_2015,ugeda_characterization_2016,sugawara_unconventional_2016}.
With respect to few-layer TMDs, experiments show that the superconducting critical temperature tends to decrease with decreasing number of layers~\cite{cao_quality_2015,xi_strongly_2015,ugeda_characterization_2016}.
The situation for the critical temperature of the CDW order is less clear, and a consensus on how it depends on the layer thickness has not yet been achieved~\cite{calandra_effect_2009,xi_strongly_2015,ugeda_characterization_2016,albertini_effect_2017}.

The metallic compound 2$H$-NbS$_2$ stands out in the TMD family, as it is superconducting with a critical temperature of $\sim$6~K, but does not exhibit CDW order in the bulk. It has recently been proposed that bulk NbS$_2$ is actually on the verge of a CDW instability~\cite{heil_origin_2017}. The strongly anharmonic phonon modes dominate the superconducting pairing in 2$H$-NbS$_2$, successfully explaining the two-gap feature observed in experiments. In Ref.~\onlinecite{heil_origin_2017} it is shown that accurate low-energy band structures and Fermi surfaces beyond density-functional theory (DFT) are crucial for achieving a better quantitative agreement with experiments. At variance with bulk NbS$_2$, the fabrication and experimental investigation of monolayer NbS$_2$ has not been reported yet.

In the present study, we want to expand on these findings and provide an extensive view on the electronic properties of bulk and monolayer NbS$_2$ beyond DFT by considering many-body perturbation theory in the framework of the $GW$ approximation~\cite{hedin_new_1965,strinati_1980,strinati_1982,hybertsen_electron_1986,giustino_gw_2010,lambert_ab_2013}. In addition, as only few $GW$ calculations have been reported for metallic TMDs~\cite{kim_quasiparticle_2017}, part of this work is to examine the various convergence parameters of the $GW$ calculation for the bulk and the monolayer system. As the main focus of this work, we compare our calculations with recent angle-resolved photoemission spectroscopy (ARPES) measurements on bulk NbS$_2$~\cite{sirica_electronic_2016}. Furthermore, we assess the effects of the increased spatial confinement when going from bulk to the monolayer, and the role of substrates. We discuss comparisons of the DFT-derived electronic structures with the quasiparticle (QP) band structures to shed light on the effects of many-body interactions.

This paper is organized as follows: In Sec.~\ref{sec:computational_details}, we provide structural details about bulk and monolayer NbS$_2$, we give a short overview of the methods used in this work, and summarize the computational setup and the convergence properties of the $GW$ calculations. In Sec.~\ref{sec:results} we discuss our results for the band structures, densities of states, and Fermi surfaces of bulk and monolayer NbS$_2$. In the case of bulk NbS$_2$ we compare our calculations with recent ARPES data. In Sec.~\ref{sec:conclusions}, we offer our conclusions and identify important avenues of future research. In the appendixes we report detailed convergence tests of the $GW$ calculations with respect to plane waves, energy cutoffs, and Brillouin-zone sampling.

\section{Computational Details}
\label{sec:computational_details}

\subsection{Structures}
\label{subsec:structures}
\begin{figure}
  \includegraphics[width=0.6\columnwidth]{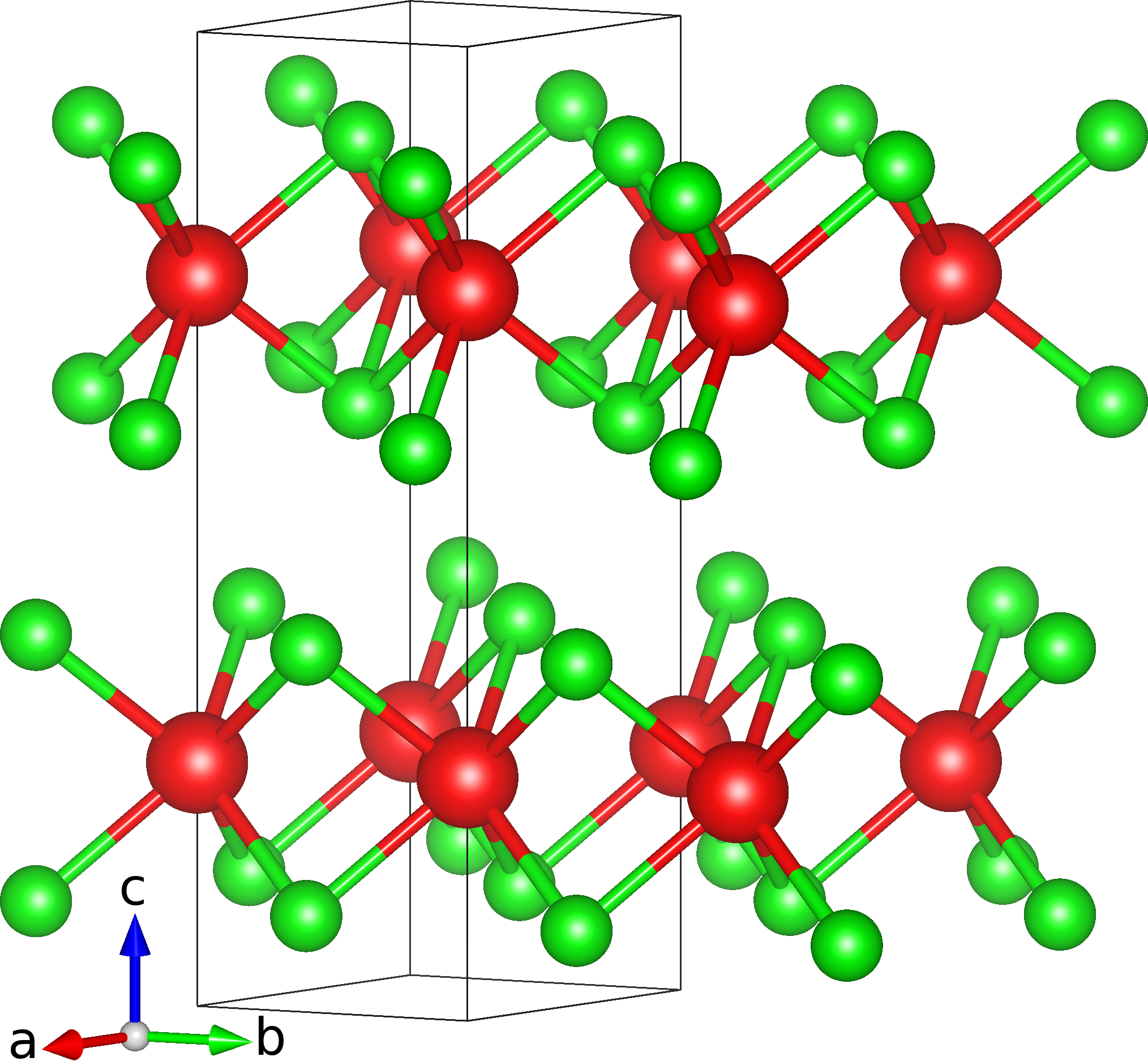}
  \caption{Ball-and-stick model of the unit cell of 2$H$-NbS$_2$. Nb atoms are shown in red, and S atoms in green.}
  \label{fig:struct}
\end{figure}

Bulk 2$H$-NbS$_2$ crystallizes in a layered hexagonal structure with space group $P6_3/mmc$. The unit cell contains two S-Nb-S layers, as shown in Fig.~\ref{fig:struct}. The Nb atoms occupy the $2b$ Wyckoff positions (0, 0, 1/4) and the S atoms are located at the $4f$ positions (1/3, 2/3, $z$). All calculations are performed in the optimized crystal structure, with lattice parameters $a=3.28$~\AA, $c/a=3.47$, and~$z=0.113$.

The NbS$_2$ monolayer is simulated by replacing one of the two S-Nb-S layers in the unit cell by a vacuum region of $10$~\AA. In order to avoid artificial effects from the nonperiodic $z$ direction in the 2D calculations, the bare Coulomb interaction is truncated along the non-periodic dimension for both the correlation and the exchange part of the self-energy~\cite{beigi06}. We checked that the changes in our results due to an increase of the vacuum region are negligible. In the monolayer, the inversion symmetry is broken, so that spin-orbit coupling (SOC) will induce a Rashba-Dresselhaus splitting~\cite{brw15,dres55} of the bands. However, the authors of Ref.~\onlinecite{yazyev_quasiparticle_2012} showed that the effects of including SOC in $GW$ calculations for a related chalcogenide are small. Hence, it should be sufficient to evaluate the magnitude of the spin splitting on the level of DFT and impose the same splitting on the $GW$ results. Kim and Son~\cite{kim_quasiparticle_2017}, for example, investigated the SOC effects along these lines in the isoelectric and isostructural NbSe$_2$. As we show in Appendix~\ref{app:soc}, for both bulk (Fig.~\ref{fig:bulk_comp_bands_dos_soc}) and monolayer NbS$_2$ (Fig.~\ref{fig:mono_comp_bands_dos_soc}) the effects of including SOC on the level of DFT is smaller than the changes in the electronic structure due to $GW$ corrections, allowing us to neglect SOC in a first approximation and solely focus on the changes of the electronic structure arising from the inclusion of many-body perturbation theory.
We also tested the influence of including van der Waals (vdW) corrections at the step of the structure optimization within DFT on our $GW$ results, and found the changes to be very small. Hence, in order to avoid any confusion, all the results in the main part of the paper are performed without vdW corrections, and a discussion on the effects of including vdW forces is provided in Appendix~\ref{app:vdW}.

\subsection{Methodology}
\label{subsec:methodology}
First-principles DFT calculations~\cite{hohenberg_inhomogeneous_1964} 
generally yield accurate predictions for ground-state properties such as structural parameters or the response to static electric and magnetic fields. The predictive power is less impressive when it comes to electronic excitations. For example, Kohn-Sham DFT tends to yield too low band gaps, as well as inaccurate band widths, densities of states, and effective masses~\cite{giustino_materials_2014}. Among the possible approaches to overcome these limitations, such as for example hybrid functionals, the quasiparticle $GW$ method has emerged as one of the most reliable tools across the materials spectrum, such as semiconductors, insulators, and metals, bulk systems, as well as surfaces and interfaces~\cite{hedin_new_1965,strinati_1980,strinati_1982,hybertsen_electron_1986}. In this method, the exchange-correlation potential $V_{xc}$ is replaced by the system's self-energy $\Sigma$, and the QP energies are obtained by performing a perturbative expansion of $\Sigma$ and the dielectric function $\epsilon$. By doing so, this method is free of the previously mentioned self-interactions and also reproduces the nonanalytic behavior when the particle number changes.\footnote{For more extensive and detailed reviews of the development and application of the $GW$ method we refer the reader to Refs.~\onlinecite{hedin_effects_1969,hybertsen_electron_1986,aryasetiawan_gw_1998,aulbur_exact-exchange-based_2000,onida_electronic_2002,giustino_gw_2010}.}
This permits us to study a wide range of diverse materials, from solids to molecules and nanostructures, as well as surfaces and interfaces~\cite{anisimov_strong_2000,faber_excited_2014}. Important applications include the correction of band gaps and bandwidths~\cite{Rasmussen_computational_2015}, and while $GW$ has traditionally been used for insulators and semiconductors, it is expected to affect the properties of metallic systems as well.

All these advantages, however, require a considerably heavier computational workload compared to DFT methods. Specifically, traditional $GW$ approaches show a very slow convergence with respect to unoccupied Kohn-Sham states~\cite{aryasetiawan_gw_1998}. To deal with these issues, Refs.~\onlinecite{giustino_gw_2010,lambert_ab_2013} developed a method to calculate the self-energy without the need to consider unoccupied states. In this approach, both the Green's function and the screened Coulomb interaction are determined by evaluating the Sternheimer equations in linear response~\cite{baroni_phonons_2001,Nguyen_Improving_2012,Pham_gw_2013}, as originally proposed in Ref.~\onlinecite{reining1997elimination} and also demonstrated in Refs.~\onlinecite{wilson2008efficient,umari2010gw}.
To this end, the Green's function $G(\mathbf{r},\mathbf{r}';\omega)$ and the screened Coulomb interaction $W(\mathbf{r},\mathbf{r}';\omega)$ are written as functions of $\mathbf{r'}$, and the space variable~$\mathbf{r}$ and the frequency $\omega$ are considered as parameters.\footnote{A more detailed derivation can be found in Ref.~\onlinecite{giustino_gw_2010} and references therein.} 
The Green's function is given by
\begin{equation}
\left( \hat{H} - \omega \right) G_{[\mathbf{r},\omega]} = - \delta_{\mathbf{r}}~,
\label{eq:G}
\end{equation}
where $\hat{H}$ represents the effective single-particle Hamiltonian. To obtain the first-order variations to the occupied electron states $\Delta \psi_{\nu [\mathbf{r},\omega]}^\pm$ corresponding to the perturbation $\Delta V_{[\mathbf{r},\omega]}(\mathbf{r}')$, one has to evaluate the two Sternheimer equations:

\begin{equation}
\left( \hat{H} - \epsilon_\nu \pm \omega \right) \Delta \psi_{\nu [\mathbf{r},\omega]}^\pm = - \left[ \theta(\epsilon_\nu) - \hat{P}_\nu \right] \Delta V_{[\mathbf{r},\omega]} \ \psi_{\nu [\mathbf{r},\omega]}^\pm~.
\label{eq:sternheimer}
\end{equation}
Here, the operator $\hat{P}_\nu$ projects onto the subspace of occupied states, with the index $\nu$ running over only occupied electron states. $\theta(\epsilon_\nu)$ accounts for the partial occupation in the case of a metallic system~\cite{Gironcoli1995}. The first-order variation within the random-phase approximation (RPA) to the single-particle density matrix $\Delta n_{[\mathbf{r},\omega]}$ reads
\begin{equation}
\Delta n_{[\mathbf{r},\omega]} = 2 \sum \limits_{\nu} \psi^*_\nu \left( \Delta \psi_{\nu [\mathbf{r},\omega]}^+ + \Delta \psi_{\nu [\mathbf{r},\omega]}^- \right)~.
\label{eq:delta_n}
\end{equation}
In Eq.~(\ref{eq:delta_n}), the prefactor $2$ takes into account the spin degeneracy and the superscript $+$~($-$) refers to the positive (negative) frequency component of the induced charge. 
In the case of the perturbation being set to the bare Coulomb interaction, i.e., $\Delta V_{[\mathbf{r},\omega]}(\mathbf{r}') = v(\mathbf{r},\mathbf{r}')$, the variation of the density matrix can be related to the dielectric matrix via~\cite{giustino_gw_2010}
\begin{equation}
\epsilon(\mathbf{r},\mathbf{r'},\omega) = \delta(\mathbf{r},\mathbf{r'})-\Delta n_{[\mathbf{r},\omega]}~,
\label{eq:epsilon}
\end{equation}
and the screened Coulomb interaction $W$ can then be obtained by inverting this matrix,
\begin{equation}
W(\mathbf{r},\mathbf{r'},\omega) = \int d\mathbf{r}'' v(\mathbf{r},\mathbf{r}'') \epsilon^{-1}(\mathbf{r}'',\mathbf{r'},\omega)~.
\label{eq:W}
\end{equation}
Another possibility to solve the Sternheimer equations would be to set the perturbation $\Delta V_{[\mathbf{r},\omega]}(\mathbf{r}') = W(\mathbf{r},\mathbf{r'},\omega)$, in which case the variation of the density yields a Hartree potential screening of the bare Coulomb interaction. The disadvantage of this approach is that Eq.~(\ref{eq:sternheimer}) then needs to be solved self-consistently.
For small systems, as for example the NbS$_2$ compounds considered in this work, the direct method outlined in Eqs.~(\ref{eq:epsilon}) and (\ref{eq:W}) is advantageous, due to the fact that the inversion of $\epsilon$ can be performed without a significant computational overhead. On the other hand, the self-consistent approach lends itself to being employed for larger systems and in cases where memory restrictions need to be met.

In short, the Sternheimer $GW$ method has the advantages (i) to be able to disregard all unoccupied states from the calculation, which often cause convergence problems in conventional $GW$ approaches~\cite{shih_quasiparticle_2010}, (ii) an improved accuracy with a similar or even smaller workflow compared to traditional $GW$ methods, and (iii) the fact that a single parameter, i.e., the kinetic-energy cutoff of the inverse dielectric matrix, controls the QP energy convergence. The above procedure also provides the complete Green's function and screened Coulomb interaction, which enables the calculation of the complete self-energy.

In the following, we will report the results of $GW$ calculations for bulk and monolayer NbS$_2$. These calculations have been performed at the $G_0W_0$ level with the {SternheimerGW} code~\cite{sternheimer_gw}. We recalculated the Fermi level using the $G_0W_0$ quasiparticle eigenvalues to ensure the number of electrons is conserved~\cite{Schindlmayr_diagrammatic_2001}. We employed the RPA for the density response, and the frequency integration has been performed using the Godby-Needs plasmon-pole approximation~\cite{godby_metal-insulator_1989}, using an imaginary pole energy of $4\;$eV. The frequency integration was performed on the imaginary axis and the self-energy on the real axis was obtained with an analytic continuation using Pad\'{e} approximants of order 11~\cite{vs77,giustino_gw_2010,lambert_ab_2013}. We employed scalar-relativistic optimized norm-conserving pseudopotentials~\cite{goedecker_separable_1996} including the semi-core electrons of Nb. We used the LDA exchange-correlation functional~\cite{ceperley1980ground} in the Perdew-Zunger (PZ) parametrization~\cite{perdew1981self} and an electronic smearing of $5\;$mRy. Our convergence studies, which we report in detail in Appendixes~\ref{app:conv_bulk} and \ref{app:conv_mono}, show that the following parameters are sufficient to describe the electronic states with an accuracy of $\sim 100 \;$meV. For bulk NbS$_2$ we employed a kinetic-energy cutoff for the plane waves of the ground-state DFT calculation of $40 \;$Ry (cf. Appendix~\ref{app:pw_bulk}), a $12 \times 12 \times 4$ $\mathbf{k}$ mesh (cf. Appendix~\ref{app:bz_bulk}), an energy cutoff of $10 \;$Ry for the dielectric matrix (cf. Appendix~\ref{app:diel_bulk}), and an energy cutoff of $25 \;$Ry for the exchange part of the self-energy (cf. Appendix~\ref{app:ex_bulk}). For monolayer NbS$_2$ we employed a kinetic-energy cutoff for the plane waves of the ground-state DFT calculation of $40 \;$Ry (cf. Appendix~\ref{app:pw_mono}), a $24 \times 24 \times 1$ $\mathbf{k}$ mesh (cf. Appendix~\ref{app:bz_mono}), an energy cutoff of $10 \;$Ry for the dielectric matrix (cf. Appendix~\ref{app:diel_mono}), and an energy cutoff of $25 \;$Ry for the exchange part of the self-energy (cf. Appendix~\ref{app:ex_mono}). 

We calculated the QP corrections to the Kohn-Sham eigenvalues for the bulk structure for a $6 \times 6 \times 2$ Brillouin-zone (BZ) grid and used maximally localized Wannier functions~\cite{marzari_maximally_2012,mostofi_updated_2014,hamann_maximally_2009} to interpolate this data onto a fine $60 \times 60 \times 20$ BZ grid. For the monolayer, we interpolated from a coarse $8 \times 8 \times 1$ BZ grid to a fine $160 \times 160 \times 1$ BZ grid. In the case of the monolayer, the Coulomb interaction has been truncated in the $z$ direction to account for the two-dimensional nature of the system.

\section{Results and Discussion}
\label{sec:results}

\subsection{Bulk NbS$_2$}
\label{subsec:bulk}
\begin{figure}
  	\includegraphics[width=1.0\columnwidth]{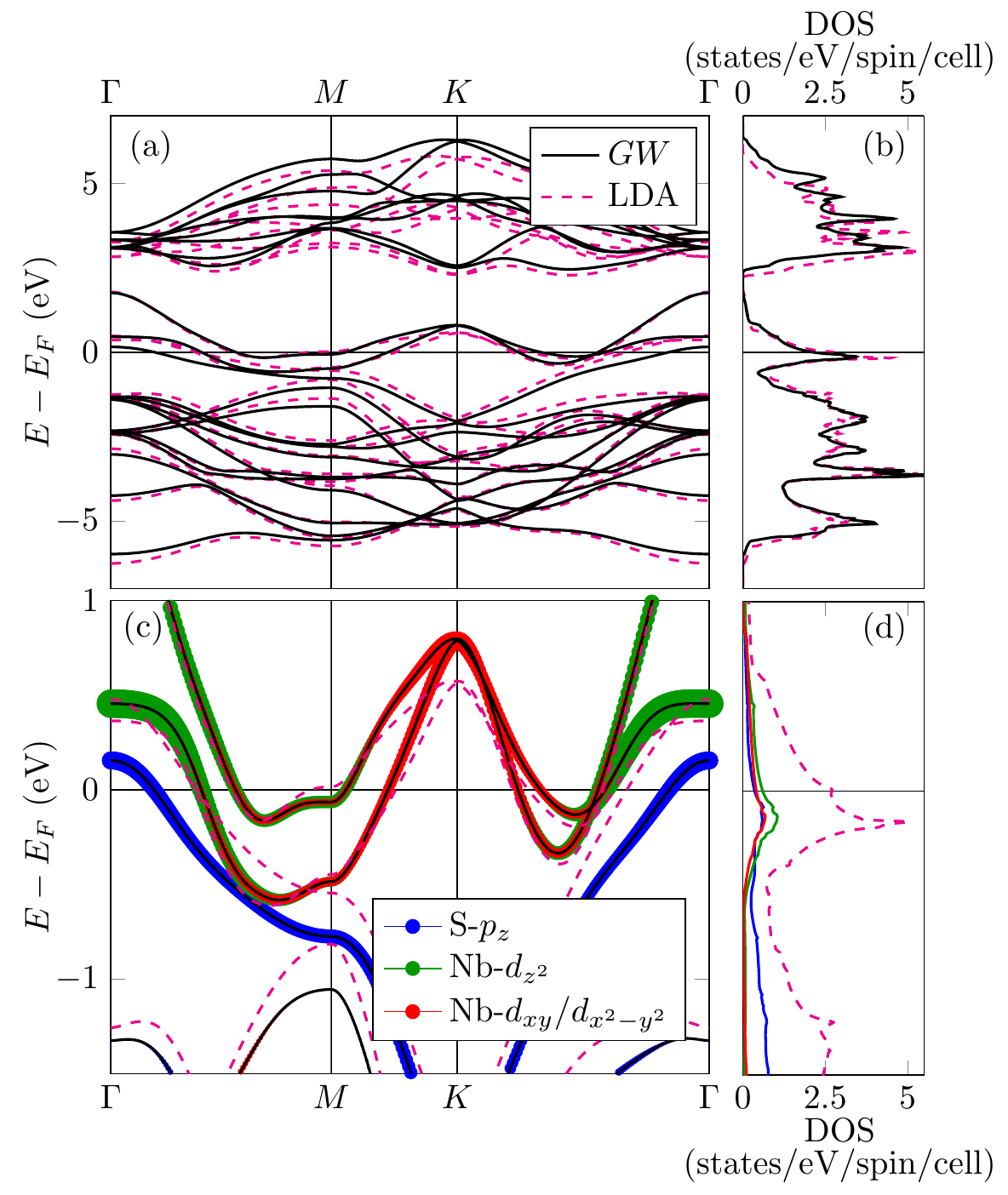}
  	\caption{Comparison of the band structure (a), (c) and DOS (b), (d) of bulk 2$H$-NbS$_2$ on the level of LDA (dashed magenta) and with $GW$ renormalization (solid black). In (c) and (d) we additionally provide the projection of the electronic states onto the S~$p_z$ (blue), the out-of-plane Nb~$d_{z^2}$ (green), and the in-plane Nb~$d_{xy/x^2-y^2}$ (red) orbitals.}
  	\label{fig:bulk_comp_bands_dos}
  \end{figure}

At first, we want to take a close look at the electronic band structures in the bulk compound. Compared to the LDA electronic structure, the inclusion of many-body interaction effects slightly increases the bandwidth of bulk 2$H$-NbS$_2$ for the electronic states up to $\pm 7$~eV around the Fermi level [cf. Fig.~\ref{fig:bulk_comp_bands_dos}(a)]. This increase originates mainly from the unoccupied bands being pushed to higher energies. We observe that at the $\Gamma$ point, the S~$p_z$ and the Nb~$d_{z^{2}}$ states are pushed apart by the inclusion of electron-electron interactions. This is in contrast to the LDA case, where they are very close in energy. The single electron band crossing the Fermi level around the $M$ point within LDA, which is of in-plane Nb~$d_{xy}/d_{x^2-y^2}$ orbital character, is renormalized to lower energies. 
We will discuss the resulting change of the topology of the Fermi surface in more detail later in this paper. For the time being we only mention that the changes to the electronic structure, mostly around the $M$ point, lead to a decrease of the DOS at the Fermi level by $18$\% [Fig.~\ref{fig:bulk_comp_bands_dos}(d)]. As is pointed out in Ref.~\onlinecite{heil_origin_2017}, this leads to a decreased electron-phonon coupling, and in turn to a lower critical superconducting temperature in this compound. At the $K$ point, we find that the unoccupied Nb~$d_{xy/x^2-y^2}$ band is pushed to higher energies. 

\begin{figure*}
	\includegraphics[width=2.07\columnwidth]{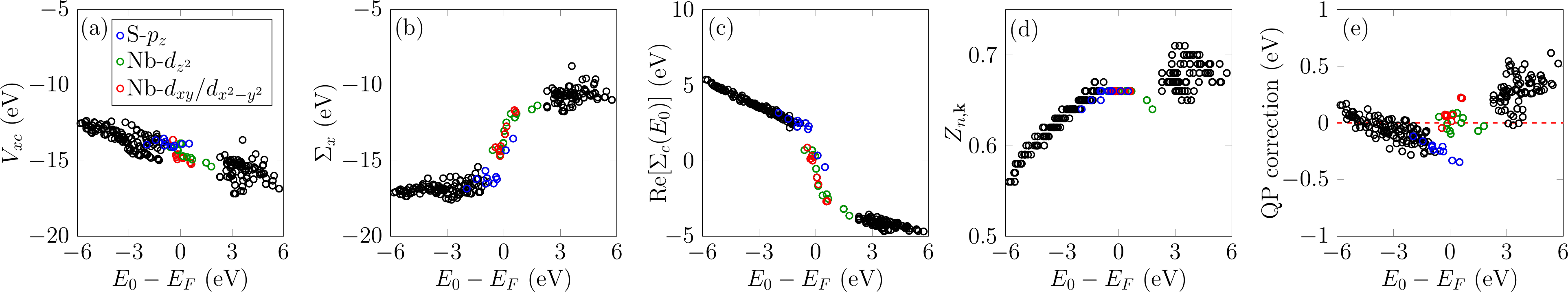}
	\caption{(a) Exchange-correlation potential $V_{xc}$, (b) exchange-part $\Sigma_{x}$ of the self-energy, (c) real part of the correlation self-energy $\Sigma_{c}$ at the DFT eigenvalues $E_0$, (d) QP renormalization factor $Z_{n,\mathbf{k}}$, and (e) QP corrections of bulk NbS$_2$ for all $\mathbf{k}$ points of the coarse Wannier grid as a function of $E_0$. The data points that belong to the Fermi-surface bands are colored according to their largest orbital character (cf. Fig.~\ref{fig:bulk_comp_bands_dos}), and the dashed red line in panel (e) serves as a guide to the eye.}
	\label{fig:bulk_Vxc_Sigmaex}
\end{figure*}

To better distinguish between the different effects the $GW$ corrections have on the DFT electronic structure, we show the exchange-correlation potential $V_{xc}$, the exchange and correlation parts of the self-energy, $\Sigma_x$ and $\Sigma_c$ respectively, as well as the QP strength $Z_{n,\mathbf{k}} = (1- \partial \text{Re} \Sigma_{n,\mathbf{k}}/\partial E)^{-1}$ and the QP correction (before the Wannier interpolation) as a function of the DFT eigenvalues in Fig.~\ref{fig:bulk_Vxc_Sigmaex}. The data points belonging to the Fermi-surface bands have been colored according to their largest orbital contribution, as detailed in Fig.~\ref{fig:bulk_comp_bands_dos}. While $V_{xc}$ is in very good approximation decreasing linearly with increasing DFT eigenvalue, $\Sigma_x$ is increasing with an abrupt change around the DFT Fermi level, as depicted in Figs.~\ref{fig:bulk_comp_bands_dos}(a) and \ref{fig:bulk_comp_bands_dos}(b) respectively. It is worth noting that $\Sigma_x$ follows the standard trend of the exchange self-energy for the homogeneous electron gas, as is typical for metals exhibiting superconducting or magnetic phases at low temperatures~\cite{mahan2013many}. The reverse trend to $\Sigma_x$ is found for ${\rm Re}\,\Sigma_{c}$, as shown in Fig.~\ref{fig:bulk_Vxc_Sigmaex}(c).

We find that in our calculations $Z_{n,\mathbf{k}}$ ranges from 0.54 to 0.71, showing that $GW$ corrections decrease the QP strength considerably. In the case that the self-energy is not explicitly $\mathbf{k}$ dependent, which is in first approximation true for the Fermi-surface bands in NbS$_2$, $Z_{n,\mathbf{k}}$ can be related to the velocity renormalization~\cite{park2007velocity} via $1-Z_{n,\mathbf{k}}^{-1} = (v_{n,\mathbf{k}}-v_{n,\mathbf{k}}^0)/v_{n,\mathbf{k}}$, with $v_{n,\mathbf{k}}$ and $v_{n,\mathbf{k}}^0$ being the interacting and noninteracting electron velocity, respectively. For bulk NbS$_2$, we find a renormalization of the electron velocity at the Fermi level of around $-51 \%$. The Fermi velocity, as well as the mass renormalization around the Fermi level and the size of the Fermi surfaces, would be experimentally accessible by quantum oscillation measurements, like the de Haas van Alphen effect~\cite{kosevich1956dehaas}. However, we are not aware of any such measurements for NbS$_2$. 

Plotting the QP corrections as function of the LDA eigenenergies, as depicted in Fig.~\ref{fig:bulk_Vxc_Sigmaex}(e), shows that the states above the Fermi level are experiencing a larger shift due to the inclusion of many-body perturbations compared to those below the Fermi level.

\begin{figure}
  	\includegraphics[width=1.0\columnwidth]{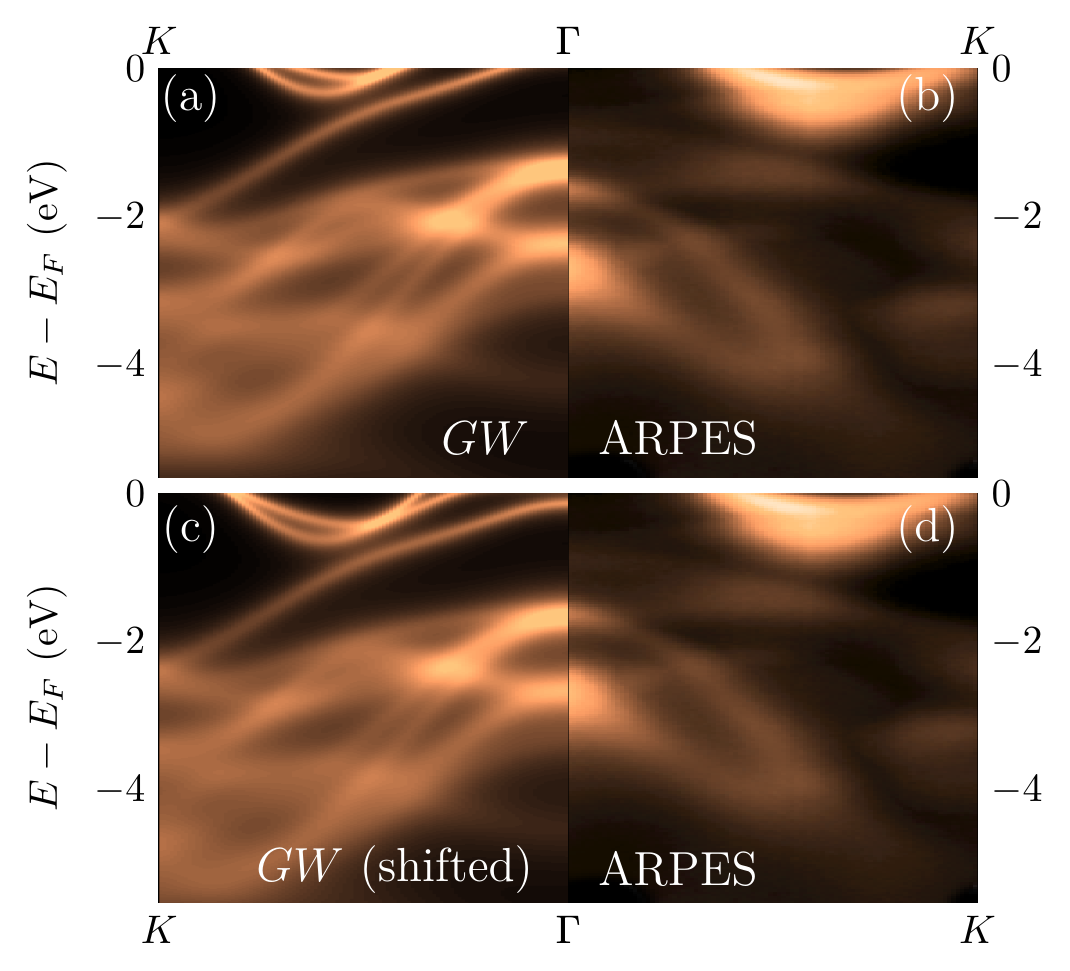}
  	\caption{Comparison of the $GW$ spectral function (a) of bulk NbS$_2$ with the ARPES measurements (b) from Ref.~\onlinecite{sirica_electronic_2016}. The spectral functions have been calculated by broadening the QP energies with a Lorentzian of energy-dependent width. (c) $GW$ spectral function shifted by $300\;$meV. (d) same as (b). Panels (b) and (d) are reprinted with permission from Ref.~\onlinecite{sirica_electronic_2016}. Copyright (2016) by the American Physical Society.}
  	\label{fig:bulk_compare_specfunc_gk_all}
  \end{figure}

In Fig.~\ref{fig:bulk_compare_specfunc_gk_all}, we compare the QP band structure with ARPES measurements from Ref.~\onlinecite{sirica_electronic_2016}. We combined the reported experimental results obtained with horizontally and vertically polarized photons in Fig.~\ref{fig:bulk_compare_specfunc_gk_all}(b). In order to account for the finite resolution in experiment, we broadened our band structure with Lorentzians with an energy-dependent full width at half maximum $\Gamma$, i.e., $\Gamma = \gamma |E-E_F| + \Gamma_0$ with $\gamma=0.15$ and $\Gamma_0=0.15$~eV, to approximately match the experimental broadening. Our calculations agree with the ARPES measurements, reproducing all the major features observed in the experiments. In particular, the two excitations slightly above and below $-2\;$eV at $\Gamma$ and their further dispersion in the direction of the $K$ point match the measurements nicely, as do the two partly occupied bands close to the Fermi level. The S $p_z$ band seems not to be visible in the ARPES data, probably due to a vanishing transition matrix element for the incident light beam. We further observe that a rigid shift of the Fermi level by $300\;$meV would bring the calculations in excellent agreement with the experiments, as shown in Fig.~\ref{fig:bulk_compare_specfunc_gk_all}(c). This energy shift corresponds to adding 0.47 electrons per formula unit (f.u.) to the system, and could be explained by intrinsic doping effects or an imperfect stoichiometry of the sample. For example, the stoichiometry Nb$_{1.09}$S$_2$ would match the observed shift of $300\;$meV. Taking into account the accuracy of our $GW$ calculations ($\sim 100\;$meV) and possible shifts in the band structure due to electron-phonon interaction ($\sim 50\;$meV) in the most favorable way, our calculations would also be in agreement with a stoichiometry of Nb$_{1.05}$S$_2$. These compositions for Nb$_{1+x}$S$_2$ are well within a range that has been realized in experiments~\cite{fisher1980stoichiometry}.
Another possible source for the observed energy shift could be due to the fact that the $G_0 W_0$ approach employed here is not a fully self-consistent method, hence the starting point of the calculation, i.e., the choice of the DFT functional, can be expected to slightly affect our results. Furthermore, as mentioned before, the $G_0 W_0$ method does not conserve the particle number and the Fermi energy has to be recomputed, leading to additional complexity~\cite{Schindlmayr_diagrammatic_2001}.

\begin{figure}
  	\includegraphics[width=1.0\columnwidth]{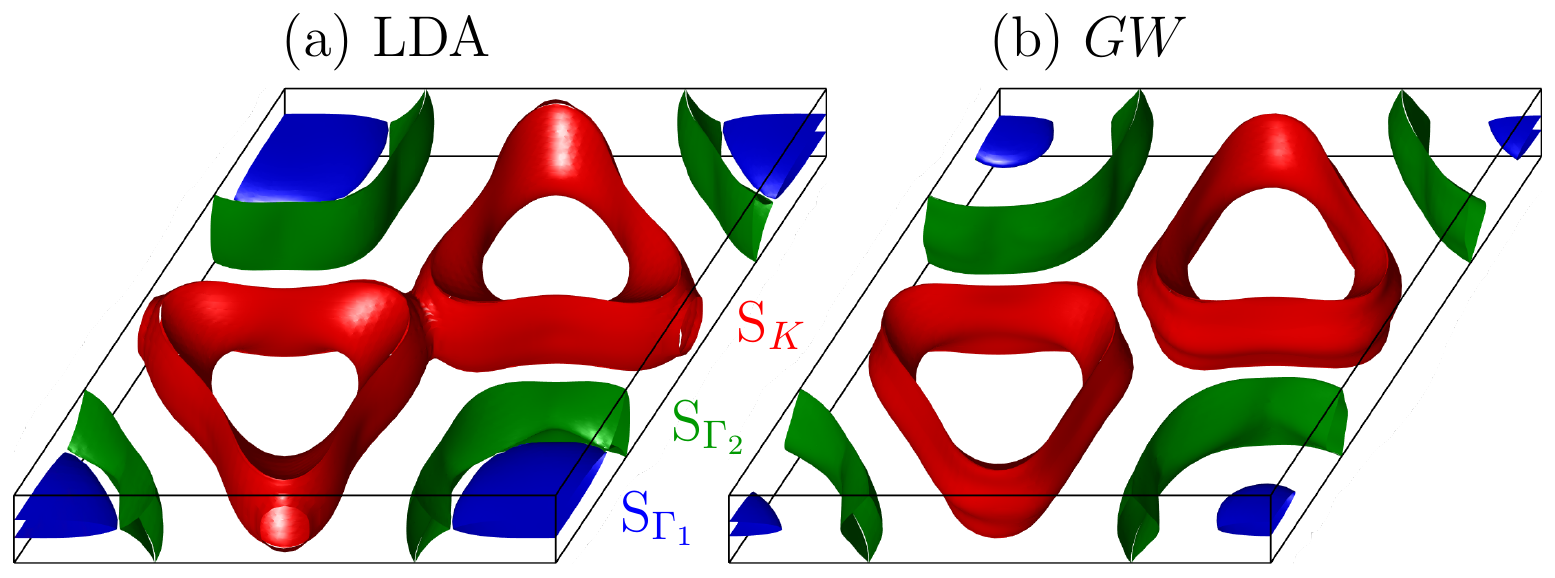}
  	\caption{Fermi surfaces of bulk NbS$_2$ calculated within LDA~(a) and $GW$~(b), where the FS $S_{\Gamma_1}$ is shown in blue, $S_{\Gamma_2}$ in green and $S_K$ in red. We want to stress that these colors do not represent the orbital contributions.}
  	\label{fig:bulk_compare_fs_arpes_1}
  \end{figure}

In the following, we want to take a detailed look at the changes of the Fermi surface, once $GW$ corrections are taken into account. In Fig.~\ref{fig:bulk_compare_fs_arpes_1} we illustrate the three sheets forming the Fermi surface: a disk-shaped pocket centered at the $\Gamma$ point ($S_{\Gamma_1}$) originating from S~$p_z$ orbitals; another $\Gamma$ centered, tube-shaped pocket ($S_{\Gamma_2}$) of Nb~$d_{z^2}$ character; and a $K$-centered, triangular-shaped pocket ($S_{K}$). The latter has Nb~$d_{z^2}$ character at the $M$ point, but changes gradually to Nb~$d_{xy/x^2-y^2}$ when moving along the surface. When electron-electron interactions are taken into account at the level of $GW$ [Fig.~\ref{fig:bulk_compare_fs_arpes_1}(b)], $S_{\Gamma_1}$ shrinks considerably, while $S_{\Gamma_2}$ becomes more rounded and grows in diameter. Another important aspect of many-body corrections is the fact that the triangular $S_{K}$ no longer extends up to the $M$ point and it is no longer connected to the neighboring $S_{K}$ Fermi surfaces. 

\begin{figure}
  	\includegraphics[width=1.0\columnwidth]{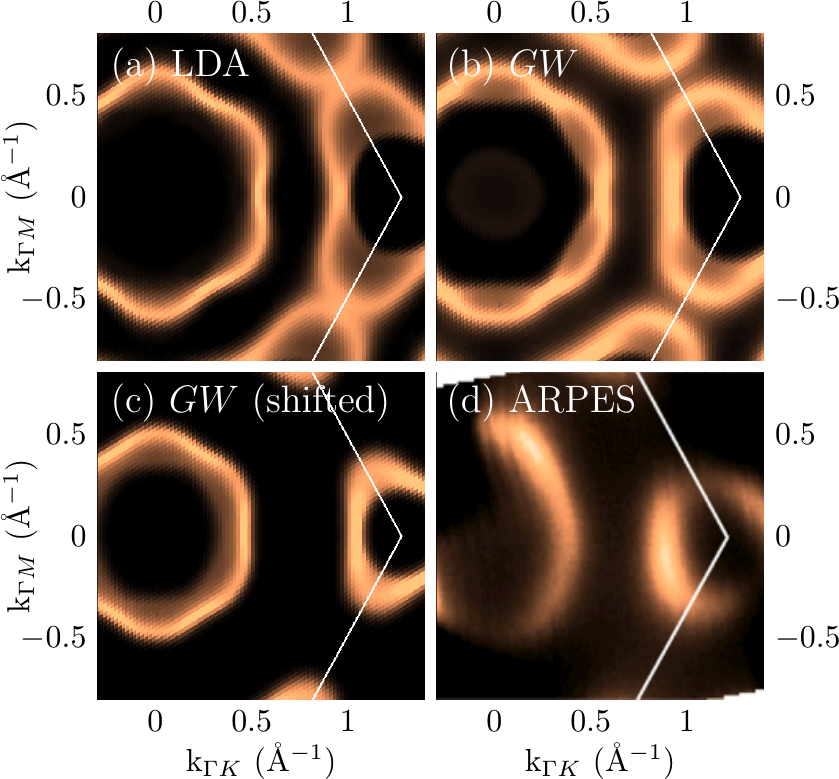}
  	\caption{Fermi surfaces of bulk NbS$_2$ calculated within LDA (a) and $GW$ (b), broadened with Lorentzians and averaged over $\mathbf{k}_z$. In panel (c), we have shifted the Fermi level by $300\;$meV, corresponding to the electronic states shown in Fig.~\ref{fig:bulk_compare_specfunc_gk_all}(c). For comparison with experiments, the measured ARPES Fermi surface from Ref.~\onlinecite{sirica_electronic_2016} is reproduced in panel~(d). Panel (d) is reprinted with permission from Ref.~\onlinecite{sirica_electronic_2016}. Copyright (2016) by the American Physical Society.}
  	\label{fig:bulk_compare_fs_arpes_2}
  \end{figure}

In order to compare our Fermi surfaces with those from experiments, we broaden the QP energies close to the Fermi level by a Lorentzian with an energy-dependent width, as detailed before, to mimic the energy and momentum resolution of ARPES measurements (Fig.~\ref{fig:bulk_compare_fs_arpes_2}). As the experiments do not resolve the out-of-plane component, we also average over $k_z$. Independent of the method, the $S_{\Gamma_1}$ Fermi surface is indiscernible from the background, in agreement with experiments~\cite{sirica_electronic_2016}, where the polarization of the incident beam probably has vanishing dipole matrix elements with the $p_z$ orbitals, as mentioned before. Only the Nb~$d$ orbitals provide visible contributions to the Fermi surface, and due to the more rounded and disconnected $S_K$ Fermi surface, the $GW$ calculation yields a better agreement with ARPES measurements than the LDA. Again, by rigidly shifting our $GW$ results by $300\;$meV, as shown in Fig.~\ref{fig:bulk_compare_fs_arpes_2}(c), we obtain very good agreement with experiments.

\subsection{Monolayer NbS$_2$}
\label{subsec:monolayer}

\begin{figure}
  	\includegraphics[width=1.0\columnwidth]{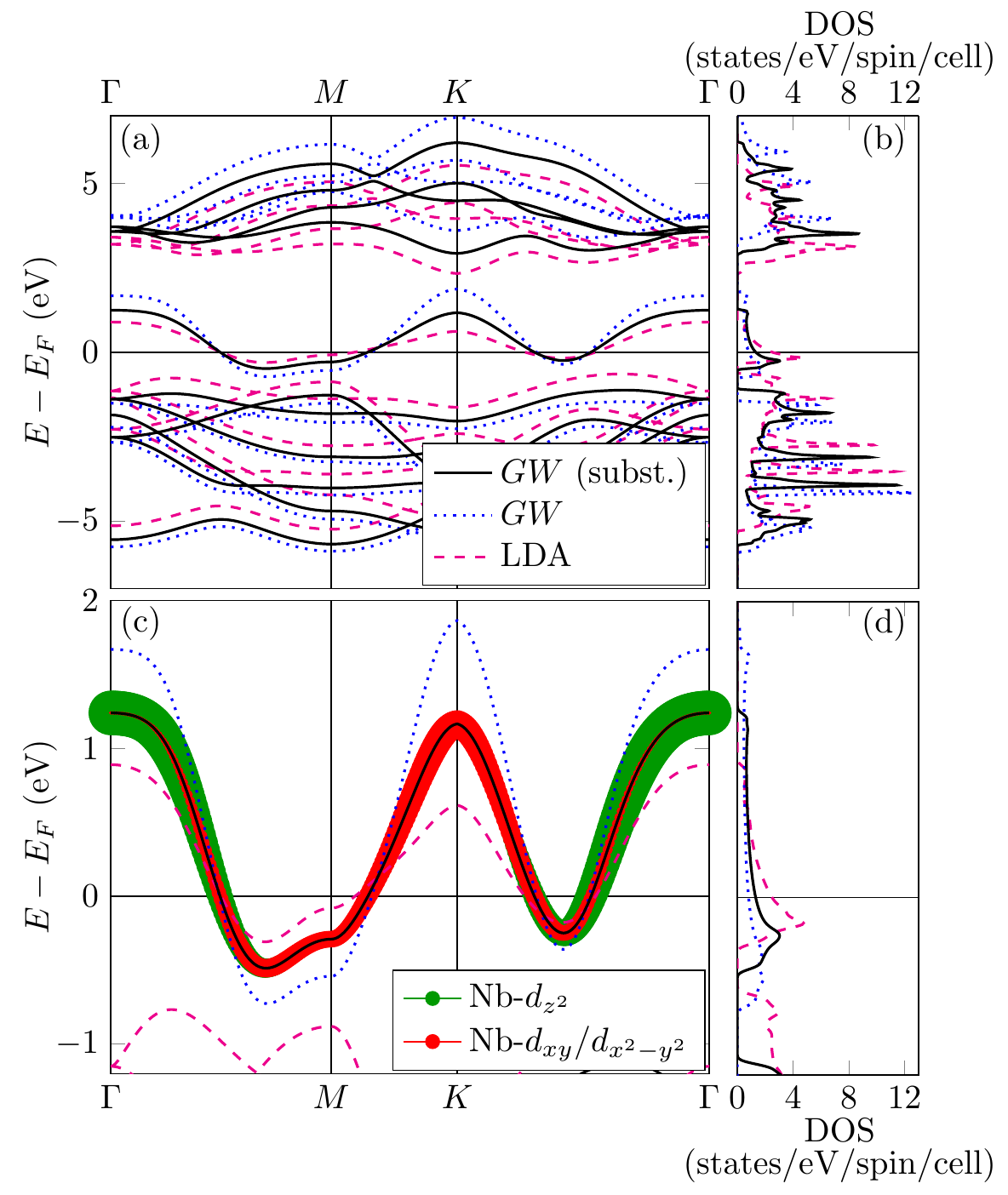}
  	\caption{Comparison of the band structure (a), (c), and DOS (b), (d) of monolayer 2$H$-NbS$_2$ on the level of LDA (dashed magenta) and with $GW$ renormalization, neglecting and including substrate effects (solid black and dotted blue lines, respectively). In (c), the orbital contributions of the out-of-plane Nb~$d_{z^2}$ and in-plane Nb~$d_{xy/x^2-y^2}$ are given by the size of the colored markers in green and red, respectively.}
  	\label{fig:mono_comp_bands_dos}
  \end{figure}

We now want to focus our attention on the monolayer of NbS$_2$. In contrast to the bulk, the monolayer is comprised of only one formula unit, and due to the fact that in the monolayer the interaction between neighboring S atoms along the $z$ direction vanishes, the bandwidth of the S~$p_z$ states decreases, and they are pushed below the Fermi energy. The Fermi surface originates therefore from only one electron band, which has out-of-plane Nb~$d_{z^2}$ character around $\Gamma$ and in-plane Nb~$d_{xy/x^2-y^2}$ character around $K$, as shown in Fig.~\ref{fig:mono_comp_bands_dos}. At the $M$ point we find a mixture of these orbital contributions. 

The impact of many-body corrections is much more pronounced in the free-standing monolayer than in the bulk system. As shown in Fig.~\ref{fig:mono_comp_bands_dos} by the dotted blue line, the band width of the electronic state crossing the Fermi level increases significantly, from $1.28\;$eV (LDA) to $2.66\;$eV ($GW$) and the DOS at the Fermi energy decreases by more than a factor of 2, i.e., from $2.48$ to $0.82\;$eV$^{-1}$.

The observed increase of $GW$ corrections in the free-standing monolayer can be attributed to the fact that the Coulomb screening is weaker in two-dimensional systems compared to their three-dimensional counterparts. For the isolated monolayer, the electric-field lines close in the vacuum region, therefore increasing the screened Coulomb interaction, as has been reported for many different systems, including BN, MoS$_2$, and graphene~\cite{park_first-principles_2009,wang_electronics_2012,kotov_electron-electron_2012,komsa_effects_2012,ramasubramaniam_large_2012,cheiwchanchamnangij_quasiparticle_2012,hwang_fermi_2012,qiu_optical_2013,ugeda_giant_2014,trolle2017model}.
This effect is particularly important when comparing calculations with experiments where few-layer systems or monolayers are placed on substrates, which will influence the screening in the sample. In order to assess the effects of the substrate, we performed an additional $GW$ calculation for the monolayer of NbS$_2$, where we incorporated this effect by considering a semi-infinite medium of dielectric constant $\epsilon_S$ on one side of the monolayer, and vacuum on the other~\cite{park_first-principles_2009}. As we are not aware of any experimental work on monolayer NbS$_2$, we chose $\epsilon_S=4$, which is a good approximation for the dielectric constants of the substrates usually used when performing measurements on monolayer NbSe$_2$, such as SiC and hexagonal BN~\cite{levinshtein_properties_2001,cao_quality_2015,xi_ising2016_2016,wang_graphene_2017,wang_high-quality_2017}.

\begin{figure*}
	\includegraphics[width=2.07\columnwidth]{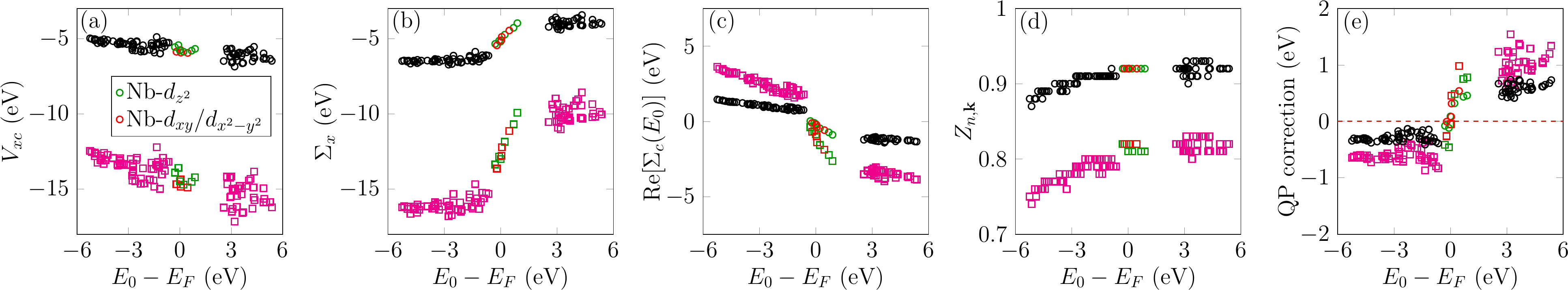}
	\caption{(a) Exchange-correlation potential $V_{xc}$, (b) exchange-part $\Sigma_{x}$ of the self-energy, (c) real part of the correlation self-energy $\Sigma_{c}$ at the DFT eigenvalues $E_0$, (d) QP renormalization factor $Z_{n,\mathbf{k}}$, and (e) QP corrections of monolayer NbS$_2$ for all $\mathbf{k}$ points of the coarse Wannier grid as a function of $E_0$ for the free-standing monolayer (squares) and with considered substrate (circles). The data points that belong to the Fermi-surface bands have been colored according to their largest orbital character (cf. Fig.~\ref{fig:mono_comp_bands_dos}), and the dashed red line in panel (e) serves as a guide to the eye.}
	\label{fig:mono_Vxc_Sigmaex}
\end{figure*}

The results of this calculation are shown as solid black lines in Fig.~\ref{fig:mono_comp_bands_dos}, where one can observe a considerable reduction of the $GW$ corrections. This becomes even more apparent when looking at $V_{xc}$, $\Sigma_{x}$, $\text{Re}[\Sigma_{c}(E_{0})]$, $Z_{n,\mathbf{k}}$, and the QP corrections, as provided in Fig.~\ref{fig:mono_Vxc_Sigmaex}, where the results of the calculations with (without) substrate are shown as circles (squares). The overall behavior of these quantities as a function of the respective DFT eigenvalues is very similar to the bulk case, i.e., $V_{xc}$ is in good approximation linearly decreasing with increasing $E_0$, while $\Sigma_{x}$ is increasing with a steep slope for values close to $E_F$, closely following the trend of a homogeneous electron gas~\cite{mahan2013many}. The reverse trend of $\Sigma_x$ is found for $\text{Re}[\Sigma_{c}(E_{0})]$. The QP strength $Z_{n,\mathbf{k}}$ is larger in both cases as compared to bulk. In particular, taking substrate effects into account leads to high QP strengths from 0.87 to 0.93, which correspond to Fermi velocity renormalizations around $-9 \%$. In contrast to the bulk, there is a marked change in the slope around the Fermi level in the $GW$ quasiparticle corrections when plotted as function of the LDA eigenvalues, as shown in Fig.~\ref{fig:mono_Vxc_Sigmaex}(e). We ascribe this behavior to the fact that the Coulomb interaction is significantly larger compared to the bulk due to the reduced electronic screening.
We want to mention at this point that our calculations for monolayer NbS$_2$ are in good agreement with those presented in Ref.~\onlinecite{kim_quasiparticle_2017} for monolayer NbSe$_2$. In general, we observe that the many-body corrections in NbS$_2$ are larger than in NbSe$_2$, but they follow the same trend, i.e., a decrease of the quasiparticle energy around the $M$ point, and an increase around $\Gamma$ and $K$. Also, the Fermi surfaces become more rounded and circular in both materials due to the inclusion of $GW$ corrections.

\begin{figure}
	\includegraphics[width=1.0\columnwidth]{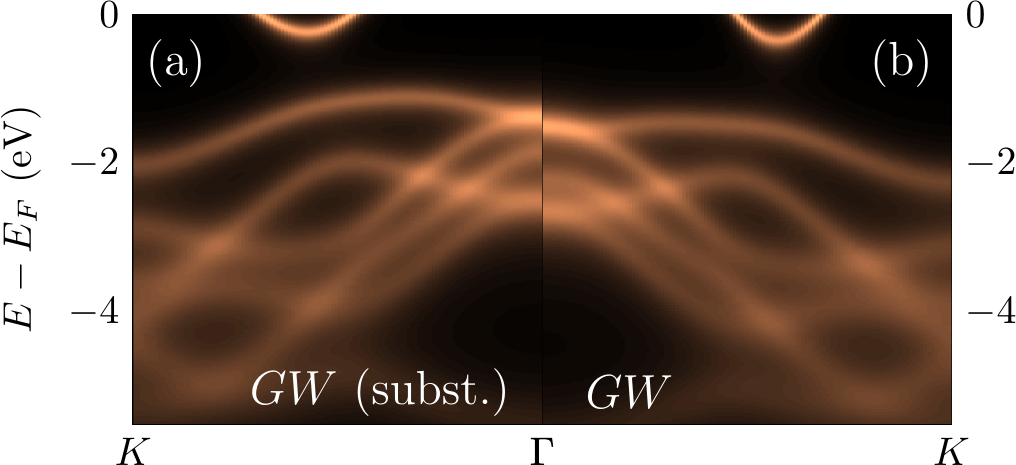}
	\caption{Spectral function for a monolayer of NbS$_2$ with (left) and without (right) substrate, along the high-symmetry $\Gamma$-$K$ path. The $GW$ quasiparticle peaks are broadened using Lorentzians of energy-dependent width.}
	\label{fig:mono_compare_specfunc_gk}
  \end{figure}

Using a Lorentzian with energy-dependent broadening to simulate the experimental resolution, we provide a prediction for the ARPES spectrum of monolayer NbS$_2$ in Fig.~\ref{fig:mono_compare_specfunc_gk}. There, we show the energy-resolved spectral function along the $\Gamma$-$K$ high-symmetry direction, and compare the results for the monolayer on a substrate and without.  The ratio of the effective mass $m_{GW}/m_\text{LDA}$ of the Fermi-surface band shown in Fig.~\ref{fig:mono_compare_specfunc_gk} increases from 0.3 to 0.5 when substrate effects are taken into account.

\begin{figure}
  	\includegraphics[width=1.0\columnwidth]{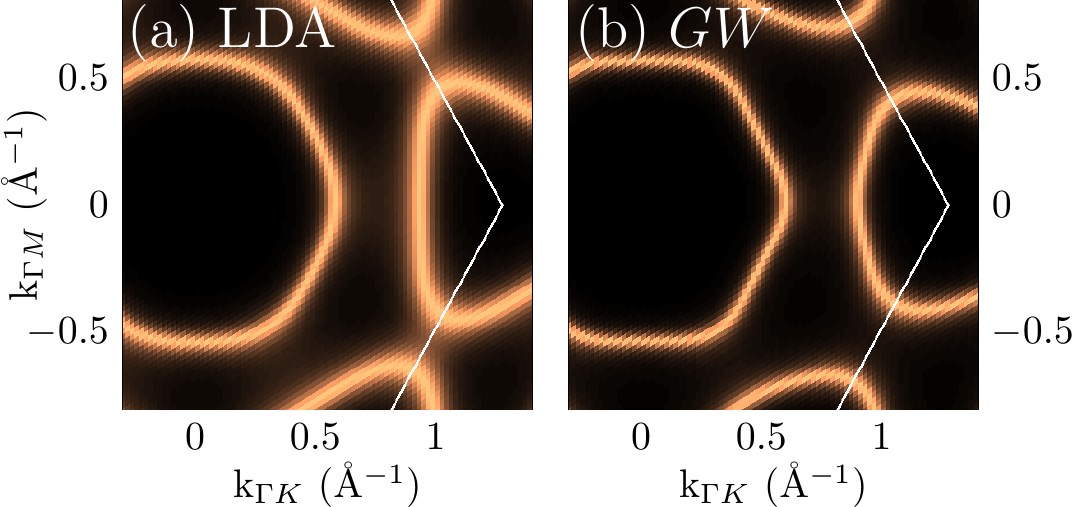}
  	\caption{Fermi surfaces for monolayer NbS$_2$, calculated within LDA (a) and $GW$ (b), where the band structure has been broadened with Lorentzians. The effects of a substrate on the $GW$ Fermi surface are negligible
     (not shown).}
  	\label{fig:mono_compare_fs_arpes}
  \end{figure}

In Fig.~\ref{fig:mono_compare_fs_arpes}, we provide our prediction for the Fermi surface of a monolayer of NbS$_2$. While the LDA Fermi surface exhibits triangular-shaped, $K$-centered sheets that are almost connected to each other, the $GW$ calculations show more circular, disconnected pockets. In general, the features of the $GW$ Fermi surface are clearer due to the steeper electron bands around the Fermi level. The Fermi surface does not change when comparing the free-standing monolayer with the monolayer on a substrate.

\section{Conclusions}
\label{sec:conclusions}
In this work, we employed the SternheimerGW method to perform a detailed investigation of the electronic properties of bulk and monolayer NbS$_2$, using many-body perturbation theory in the framework of the $GW$ approximation. We document in the appendixes convergence studies for the $GW$ calculations, to serve as a solid technical foundation for future investigations including many-body corrections in metallic transition-metal dichalcogenides. 

We analyzed the individual components of the self-energy, and found that $\Sigma_x$ behaves as in the the homogeneous electron gas. The QP strength is found to be between 0.54 and 0.74 for the bulk, and between 0.87 and 0.93 for the monolayer (when we take into account substrate screening).

For the bulk system, we find that the inclusion of many-body corrections increases the bandwidth of the S~$p$ and Nb~$d$ bands by pushing the unoccupied states to higher energies, and a decrease of the DOS at the Fermi level by $18 \%$. This decrease of the DOS has important effects on the electron-phonon coupling and superconductivity in this compound. In addition, we have compared our calculations with ARPES measurements and found that including many-body effects improves the agreement between theory and experiments in terms of the shape of the Fermi surface and the low-energy electronic structure. We also observe that a Fermi-level shift of $~300\;$meV brings our calculations in very good agreement with the ARPES measurements. We proposed that such a shift could arise from unintentional doping in the experimental sample.

For the monolayer system, we observe a strong increase of the screened Coulomb interaction as compared to bulk due to its two-dimensional nature, and we have shown that the consideration of a substrate via changing the dielectric constant on one side of the monolayer results in a marked reduction of the interaction potential and the effects of the many-body corrections. When we take into account the screening from the substrate, the electronic bandwidth is smaller than in the free-standing case, and the DOS at the Fermi level is larger. This in turn also means that the electron-phonon coupling and superconducting properties will be sensitive to the choice of substrate. As another example of the importance of substrate effects, we find that the ratio of the effective masses is 1.7 times larger for the NbS$_2$ monolayer on a substrate compared to the free-standing monolayer. As we are not aware of any experimental work on monolayer NbS$_2$, these calculations provide a first insight into the electronic properties of this compound beyond DFT.

In summary, this work highlights the importance to include many-body perturbation theory corrections in the study of metallic TMDs. These corrections will lead to improved carrier velocities and densities of states, and are crucial for a more accurate understanding of charge transport and superconductivity in these compounds.

\begin{acknowledgments}
We are grateful to Professor N. Mannella and the other authors of Ref.~\onlinecite{sirica_electronic_2016} for the permission to use their ARPES results for comparison with our data, and we acknowledge fruitful discussions with G. Volonakis and N. Zibouche.
This work was supported by the Austrian Science Fund (FWF) Project No. J 3806-N36, 
the Leverhulme Trust (Grant No. RL-2012-001), the UK Engineering and Physical Sciences Research Council (Grant No. EP/M020517/1), the Graphene Flagship (Horizon 2020 Grant No. 785219 - GrapheneCore2), the University of Oxford Advanced Research Computing (ARC) facility (http://dx.doi.org/810.5281/zenodo.22558), the ARCHER UK National Supercomputing Service under the `AMSEC' Leadership project, the Vienna Scientific Cluster (VSC), and the Cambridge Service for Data Driven
Discovery (CSD3) funded by EPSRC (grant EP/P020259/1). We further acknowledge PRACE for awarding us access to Cartesius at SURFsara, Netherlands; Abel at UiO, Norway, and MareNostrum at BSC-CNS, Spain.
\end{acknowledgments}

\appendix
\section{Effect of SOC}
\label{app:soc}

\begin{figure}
	\includegraphics[width=0.9\columnwidth]{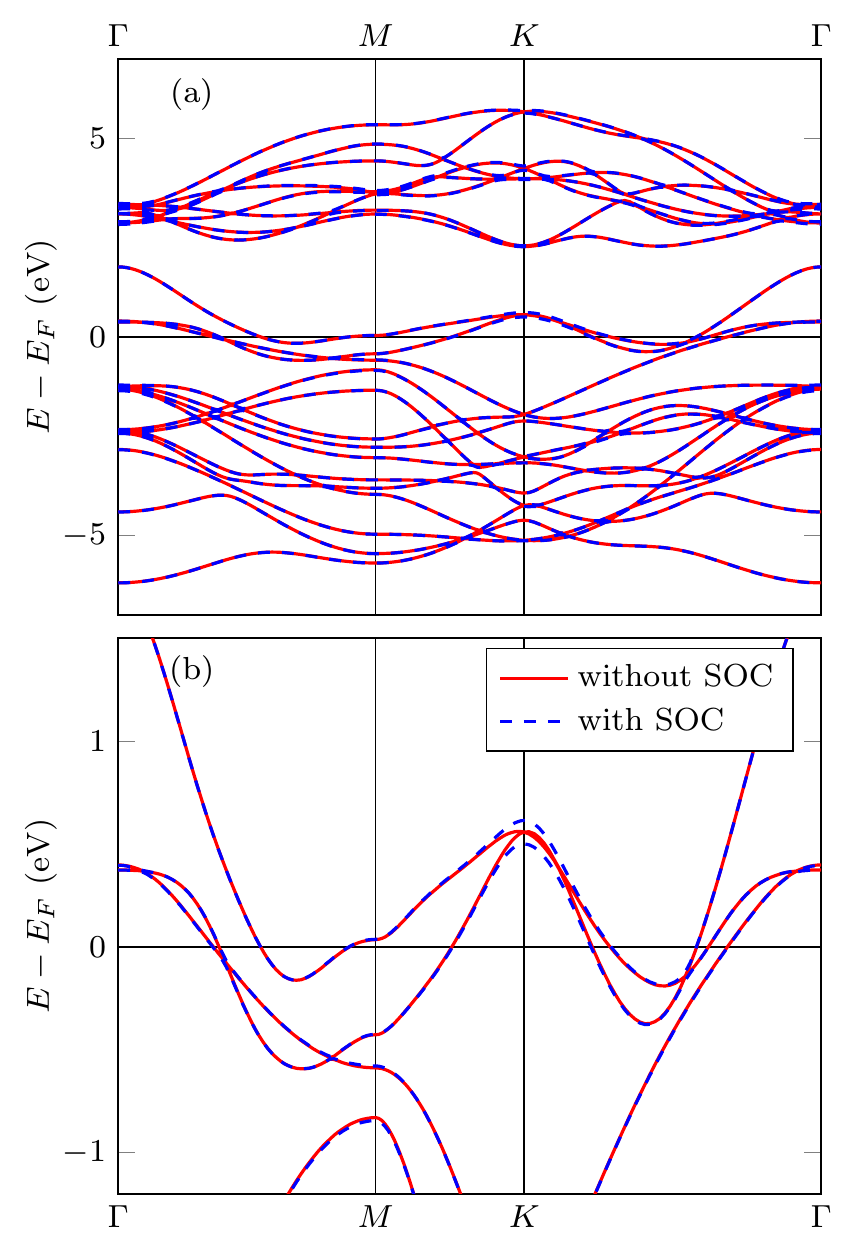}
	\caption{Comparison of the LDA band structure of bulk NbS$_2$ with (dashed, blue) and without (solid red) SOC. Panel (b) is a zoom of (a).}
	\label{fig:bulk_comp_bands_dos_soc}
\end{figure}

\begin{figure}
	\includegraphics[width=0.9\columnwidth]{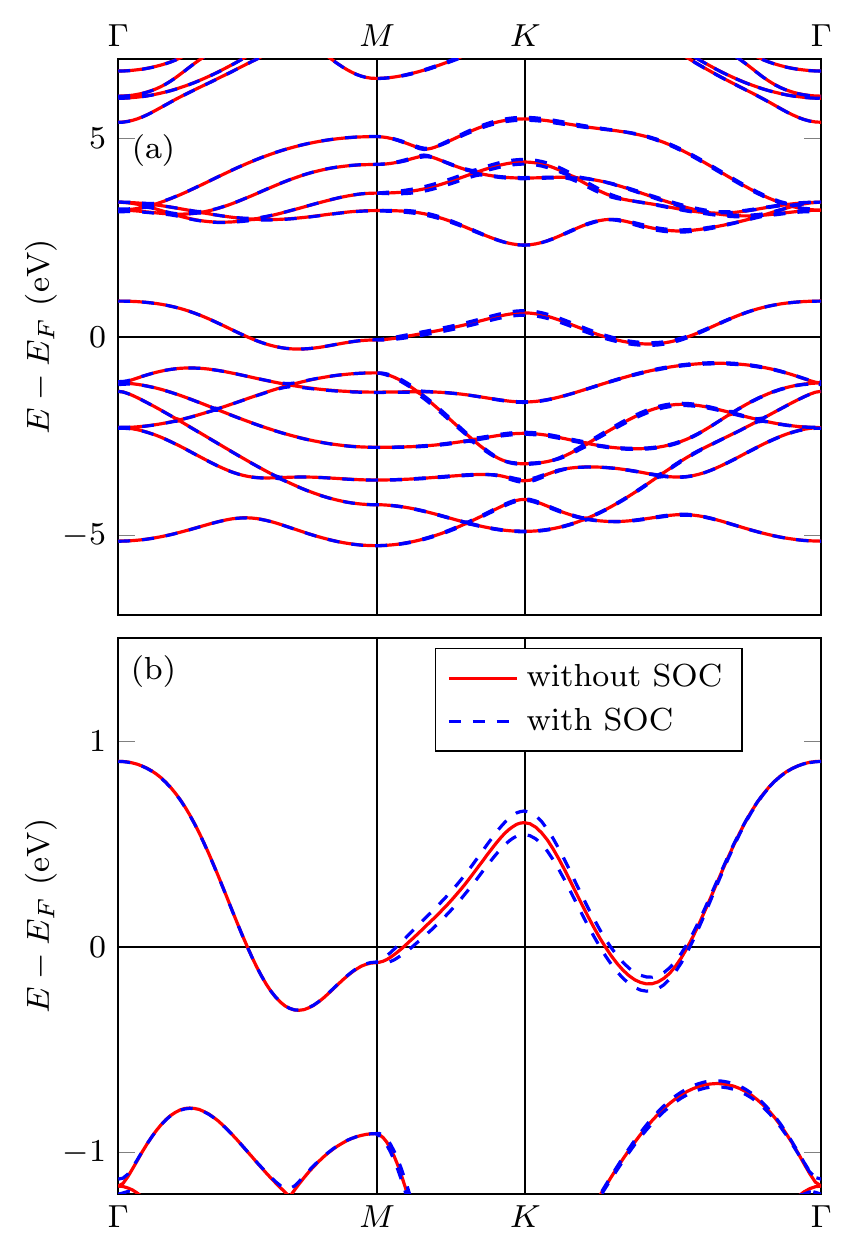}
	\caption{Comparison of the LDA band structure of the free-standing NbS$_2$ monolayer with (dashed, blue) and without (solid red) SOC. Panel (b) is a zoom of (a).}
	\label{fig:mono_comp_bands_dos_soc}
\end{figure}

In Figs.~\ref{fig:bulk_comp_bands_dos_soc} and \ref{fig:mono_comp_bands_dos_soc} we provide a comparison of the DFT band structure with (dashed blue) and without (solid red) SOC for bulk and monolayer NbS$_2$, respectively. For these calculations, fully relativistic optimized norm-conserving Vanderbilt pseudopotentials~\cite{hamann_optimized_2013,schlipf_optimization_2015} within PZ~\cite{perdew1981self} that include the semi-core electrons of Nb have been used. As one can see, SOC leads to band splittings around the $K$ point. For the bulk, we calculate a splitting of $\sim$74~meV and for the free-standing monolayer a splitting of $\sim$52~meV. These values are within the accuracy of our $GW$ results, and are therefore neglected.

\section{Effect of van der Waals corrections}
\label{app:vdW}

\begin{figure}
    \includegraphics[width=1.0\columnwidth]{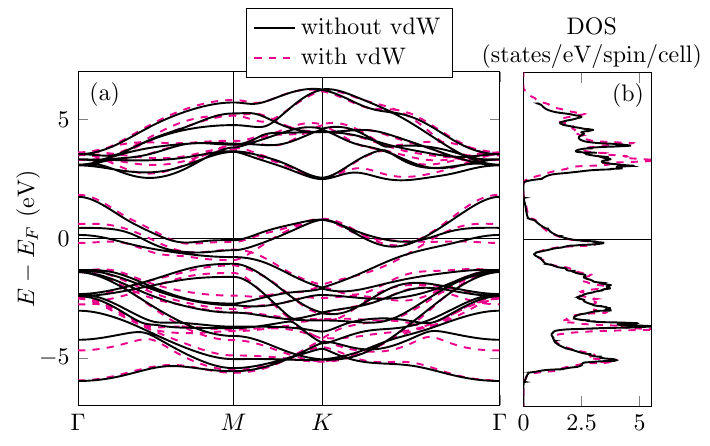}
    \caption{Band structure of bulk NbS$_2$ with included $GW$ corrections, where the structure has been optimized at the DFT level with (dashed, magenta) and without (black, solid) van der Waals corrections.}
    \label{fig:bulk_comp_bands_dos_vdW}
\end{figure}

In this appendix we discuss the effects of including van der Waals corrections on the band structure of bulk NbS$_2$. In Fig.~\ref{fig:bulk_comp_bands_dos_vdW}, we show as solid black lines the band structure at the level of $GW$ without considering vdW effects, while for the dashed magenta line we have included vdW interactions according to the semiempirical Grimme D2 corrections~\cite{grimme_semiempirical_2006}. As can be appreciated in this figure, the changes to the band structure are small, and the DOS in the low-energy range is almost identical. The most marked difference is the fact that with vdW corrections, the S~$p_z$ close to the $\Gamma$ point falls below the Fermi energy.

\begin{figure}
    \includegraphics[width=1.0\columnwidth]{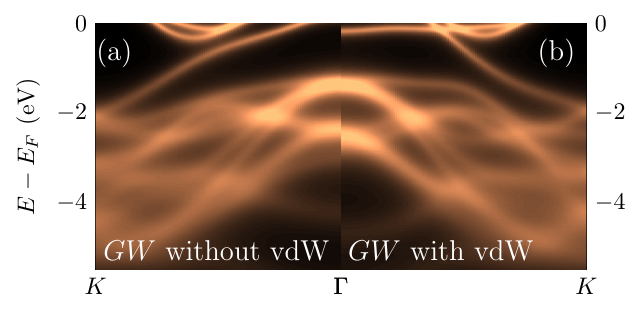}
  	\caption{Comparison of the $GW$ spectral function of bulk NbS$_2$ (a) without and (b) with vdW corrections during the structure optimization. The spectral functions have been calculated by broadening the QP energies with a Lorentzian of energy-dependent width, as in Fig.~\ref{fig:bulk_compare_specfunc_gk_all}.}
\label{fig:bulk_compare_specfunc_gk_all_vdW}
\end{figure}

This can also be observed in Fig.~\ref{fig:bulk_compare_specfunc_gk_all_vdW}, where we show the $GW$ spectral function with and without vdW corrections during the structure optimization. Apart from the QP band coming from S~$p_z$ states, the two spectral functions are almost indistinguishable.

\begin{figure}
    \includegraphics[width=1.0\columnwidth]{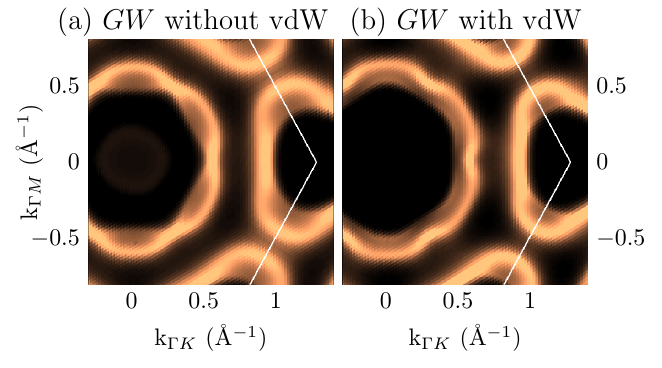}
    \caption{$GW$ Fermi surfaces of bulk NbS$_2$ calculated (a) without and (b) with van der Waals corrections during the structure optimization. The spectral functions have been calculated by broadening the QP energies with a Lorentzian of energy-dependent width, as in Fig.~\ref{fig:bulk_compare_fs_arpes_2}.}
    \label{fig:bulk_compare_fs_arpes_vdW}
\end{figure}

Moving on to the Fermi surfaces shown in Fig.~\ref{fig:bulk_compare_fs_arpes_vdW}, we find that, apart from very small, the structure optimized with vdW corrections leads to very similar results for the $GW$ spectral function at low energies.

\section{Convergence studies for bulk NbS$_2$}
\label{app:conv_bulk}

\subsection{Plane wave cutoff for ground state DFT calculations}
\label{app:pw_bulk}
First, we test the convergence of our $GW$ calculations with respect to the kinetic-energy cutoff $E_k$ for the plane waves used in the ground-state DFT calculation. We report the results of this study in Table~\ref{tab:bulk_inveps_vs_scfecut}, where we show the inverse of the diagonal element of the dielectric matrix $\epsilon_i = \epsilon_{\mathbf{G}_i,\mathbf{G}_i} (\mathbf{q},\omega)$ with $\mathbf{G}_i = (i,0,0) 2\pi/a$, evaluated at $\mathbf{q}=\Gamma$ and $\omega=0$.

In Table~\ref{tab:bulk_Sex_vs_scfecut}, we show the values of $\Sigma_x$ of the four QP states at $\Gamma$ closest to the Fermi level as a function of the cutoff value. As one can see, our target accuracy of $<100 \;$meV is already met for $30\;$Ry. The maximum deviation falls below $30\;$meV for a cutoff of $40\;$Ry.

\begin{table}[h!]
	\begin{center}
		\caption{The inverse of $\epsilon_i = \epsilon_{\mathbf{G}_i,\mathbf{G}_i} (\mathbf{q},\omega)$ with $\mathbf{G}_i = (i,0,0) 2\pi/a$, evaluated at $\mathbf{q}=\Gamma$ and $\omega=0$, as a function of the kinetic-energy cutoff $E_k$ for the plane waves of the DFT ground-state calculation. For these calculations, we used an $8 \times 8 \times 4$ $\mathbf{k}$ grid, an energy cutoff in the dielectric matrix of $10\;$Ry, and an energy cutoff of $20\;$Ry in the exchange part of the self-energy.}
		\label{tab:bulk_inveps_vs_scfecut}
		\begin{tabular}{c|c|c|c} 
			$E_k$ (Ry) & $1/\epsilon_2$ & $1/\epsilon_3$ & $1/\epsilon_4$ \\ \hline
			30 & 0.049 & 0.192 & 0.335 \\ \hline
			40 & 0.045 & 0.188 & 0.328 \\ \hline
			50 & 0.046 & 0.188 & 0.329 \\ \hline
			60 & 0.046 & 0.188 & 0.329 \\ \hline
			70 & 0.046 & 0.188 & 0.329 \\ \hline
		\end{tabular}
	\end{center}
\end{table}

\begin{table}[h!]
	\begin{center}
		\caption{$\Sigma_x (\mathbf{k})$ of the four electronic states closest to the Fermi energy at $\mathbf{k}=\Gamma$ (at the DFT level) as a function of the kinetic-energy cutoff $E_k$ for the plane waves of the DFT ground-state calculation. For these calculations, we used an $8 \times 8 \times 4$ $\mathbf{k}$ grid, an energy cutoff in the dielectric matrix of $10\;$Ry, and an energy cutoff of $20\;$Ry in the exchange part of the self-energy.}
		\label{tab:bulk_Sex_vs_scfecut}
		\begin{tabular}{c|c|c|c|c} 
			$E_k$ (Ry) & band 1 (eV) & band 2 (eV) & band 3 (eV) & band 4 (eV) \\ \hline
			30 & -16.83 & -11.61 & -13.70 & -10.77  \\ \hline
			40 & -16.82 & -11.55 & -13.73 & -10.72  \\ \hline
			50 & -16.82 & -11.52 & -13.75 & -10.71  \\ \hline
			60 & -16.81 & -11.51 & -13.75 & -10.70  \\ \hline
			70 & -16.82 & -11.51 & -13.75 & -10.70  \\ \hline
		\end{tabular}
	\end{center}
\end{table}

\subsection{Brillouin zone sampling}
\label{app:bz_bulk}
\begin{figure}
  	\includegraphics[width=1.0\columnwidth]{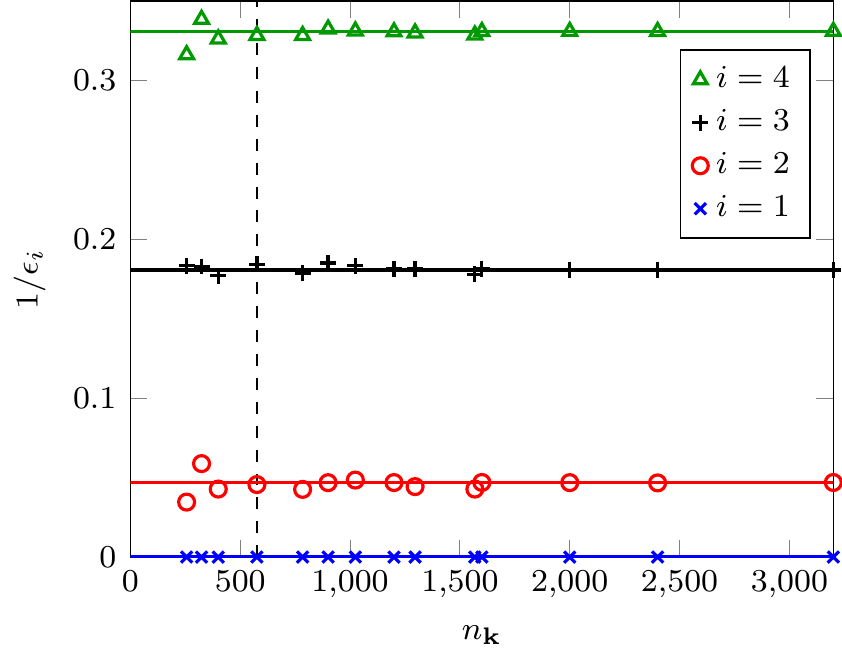}
  	\caption{The inverse of the first four diagonal elements of $\epsilon_i = \epsilon_{\mathbf{G}_i,\mathbf{G}_i} (\mathbf{q},\omega)$ with $\mathbf{G}_i = (i,0,0) 2\pi/a$, evaluated at $\mathbf{q}=\Gamma$ and $\omega=0$, as a function of the number of $\mathbf{k}$ points. The vertical dashed line indicates the value for the $12 \times 12 \times 4$ BZ grid used in the final calculations.}
  	\label{fig:bulk_inveps_vs_nk}
  \end{figure}

In {SternheimerGW}, there are three BZ grids to consider: (i) the grid used to sample the exchange, (ii) the grid used to sample the correlation, and (iii) the grid used to sample the dielectric response (in our calculations, this grid is set to be the same as for the exchange). The dependence of the correlation part of the self-energy on the number of points used to sample the Brillouin zone is shown in Fig.~\ref{fig:bulk_inveps_vs_nk}, where we plot the inverse of the first four diagonal elements of $\epsilon_i = \epsilon_{\mathbf{G}_i,\mathbf{G}_i} (\mathbf{q},\omega)$ with $\mathbf{G}_i = (i,0,0) 2\pi/a$ evaluated at $\mathbf{q}=\Gamma$ and $\omega=0$.

\begin{figure}
  	\includegraphics[width=1.0\columnwidth]{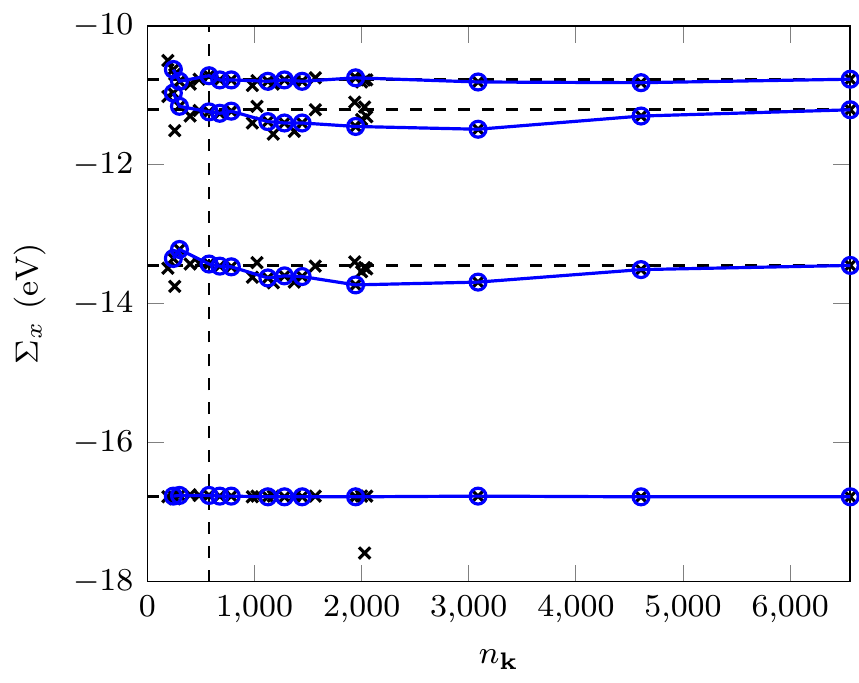}
  	\caption{$\Sigma_x$ at $\Gamma$ of the four electronic states closest to the Fermi energy (at DFT level) as a function of the number of BZ grid points. The vertical dashed line indicates the value for the $12 \times 12 \times 4$ $\mathbf{k}$ grid used in the final calculations. Data points that have a common ratio $n_{\mathbf{k}_{x,y}}/n_{\mathbf{k}_{z}}=3$ are connected with a blue line.}
  	\label{fig:bulk_Sx_vs_nk}
  \end{figure}

While the convergence of the correlation part is quite smooth, the exchange part of the self-energy is more difficult to converge with respect to the number of BZ points. We notice that achieving convergence for the two bands closest to the Fermi level is particularly difficult, while the convergence of the other states is very fast. In Fig.~\ref{fig:bulk_Sx_vs_nk} we show $\Sigma_x$ at $\Gamma$ as a function of the number of BZ points $n_{\mathbf{k}}$, where we have connected with a blue line those points that have a common $n_{\mathbf{k}_{x,y}}/n_{\mathbf{k}_z} = 3$ ratio, similar to the lattice vector lengths' ratio. The vertical dashed lines indicate the values for a $27 \times 27 \times 9$ BZ grid, which we consider converged. The large deviations between single points and the bulging downwards of the blue curve are largely due to finite-size effects coming from the $\mathbf{k}_z$ sampling. As calculations for a $27 \times 27 \times 9$ grid are at the moment not feasible, we chose a $12 \times 12 \times 4$ grid that allows for reasonably fast calculations while still maintaining an accuracy of $<100\;$meV relative to the $27 \times 27 \times 9$ grid.

\subsection{Energy cutoff $E_c$ for the correlation part of the self-energy $\Sigma_c$}
\label{app:diel_bulk}
In Table~\ref{tab:bulk_W_vs_ecutsco} we report the convergence of the diagonal elements of the screened Coulomb interaction matrix $W_{i} = W_{\mathbf{G}_i,\mathbf{G}_i}(\mathbf{q},\omega)$ with $\mathbf{G}_i = (i,0,0) 2\pi/a$, evaluated at $\mathbf{q}=\Gamma$ and $\omega=0$, as a function of the energy cutoff $E_c$ of the dielectric matrix.

\begin{table}[h!]
	\begin{center}
		\caption{Diagonal elements of the screened Coulomb interaction $W_{i} = W_{\mathbf{G}_i,\mathbf{G}_i}(\mathbf{q},\omega)$ with $\mathbf{G}_i = (i,0,0) 2\pi/a$, evaluated at $\mathbf{q}=\Gamma$ and $\omega=0$, as a function of the energy cutoff $E_c$ for the dielectric matrix. For these calculations, we used an $8 \times 8 \times 4$ $\mathbf{k}$ grid and a plane-wave cutoff of $40\;$Ry.}
		\label{tab:bulk_W_vs_ecutsco}
		\begin{tabular}{c|c|c|c|c}
			$E_c$ (Ry) & $W_{2}$     & $W_{4}$     & $W_{10}$   & $W_{14}$ \\ \hline
			 6         & -0.9013     &  -0.6964    &  -0.5344    & -0.5071     \\ \hline
			 8         & -0.9010     &  -0.6961    &  -0.5337    & -0.5066     \\ \hline
			10         & -0.9008     &  -0.6950    &  -0.5334    & -0.5065     \\ \hline
			12         & -0.9006     &  -0.6944    &  -0.5333    & -0.5064     \\ \hline
			14         & -0.9006     &  -0.6940    &  -0.5335    & -0.5063     \\ \hline
			16         & -0.9005     &  -0.6936    &  -0.5331    & -0.5064     \\ \hline
		\end{tabular}
	\end{center}
\end{table}

In Table~\ref{tab:bulk_bands_vs_ecutsco} we show the convergence behavior of the $GW$ quasiparticle energies at $\Gamma$ closest to the DFT Fermi energy as a function of the energy cutoff $E_c$ for the dielectric matrix. The convergence is relatively fast, allowing us to choose a value of $10\;$Ry for all subsequent calculations

\begin{table}[h!]
    \begin{center}
        \caption{Quasiparticle eigenvalues of the four electronic states at $\Gamma$ closest to the Fermi energy (at DFT level) as a function of the energy cutoff $E_c$ for the dielectric matrix. For these calculations, we used an $8 \times 8 \times 4$ $\mathbf{k}$ grid, a plane-wave cutoff of $40\;$Ry, and an energy cutoff for the exchange part of 20~Ry.}
        \label{tab:bulk_bands_vs_ecutsco}
        \begin{tabular}{c|c|c|c|c}
            $E_c$ (Ry) & band 1 (eV) & band 2 (eV) & band 3 (eV) & band 4 (eV) \\ \hline
            6          &  7.67       & 14.26       & 11.61       &  16.56      \\ \hline
            8          &  8.28       &  9.47       & 10.78       &  12.00      \\ \hline
            10         &  8.27       &  9.43       & 10.56       &  11.70      \\ \hline
            12         &  8.24       &  9.39       & 10.40       &  11.54      \\ \hline
            14         &  8.20       &  9.36       & 10.31       &  11.44      \\ \hline
            16         &  8.19       &  9.34       & 10.26       &  11.41      \\ \hline
        \end{tabular}
    \end{center}
\end{table}

\subsection{Energy cutoff $E_x$ for the exchange part of the self-energy $\Sigma_x$}
\label{app:ex_bulk}
In order to test the dependence of our results on the energy cutoff $E_x$ of the exchange part of the self-energy, we again inspected its values for electronic states closest to the Fermi level at the $\Gamma$ point (see Table~\ref{tab:bulk_Sex_vs_ecutsex}). While the deviations between calculations with cutoffs of $20$ and $35 \;$Ry are larger than $100 \;$meV for three of the bands, these discrepancies are reduced to $<50 \;$meV if a cutoff of $25 \;$Ry is chosen.

\begin{table}[h!]
	\begin{center}
		\caption{$\Sigma_x (\mathbf{k})$ of the four electronic states at $\Gamma$ closest to the Fermi energy (at DFT level) as a function of the energy cutoff $E_x$. For these calculations, we used an $8 \times 8 \times 4$ $\mathbf{k}$ grid, an energy cutoff in the dielectric matrix of $10 \;$Ry, and a plane-wave cutoff of $40 \;$Ry.}
		\label{tab:bulk_Sex_vs_ecutsex}
		\begin{tabular}{c|c|c|c|c} 
			$E_x$ & band 1 (eV) & band 2 (eV) & band 3 (eV) & band 4 (eV) \\ \hline
			20 	  & -16.81  & -11.51  & -13.75  & -10.70 \\ \hline
			25 	  & -16.88  & -11.61  & -13.77  & -10.80 \\ \hline
			30 	  & -16.90  & -11.65  & -13.78  & -10.84 \\ \hline
			35 	  & -16.91  & -11.66  & -13.78  & -10.85 \\ \hline
		\end{tabular}
	\end{center}
\end{table}

\section{Convergence studies for monolayer NbS$_2$}
\label{app:conv_mono}

\subsection{Plane wave cutoff for ground state DFT calculations}
\label{app:pw_mono}
In Table~\ref{tab:mono_W_vs_scfecut} we report the convergence of the diagonal elements of the screened Coulomb interaction $W_{i} = W_{\mathbf{G}_i,\mathbf{G}_i}(\mathbf{q},\omega)$ with $\mathbf{G}_i = (i,0,0) 2\pi/a$, evaluated at $\mathbf{q}=\Gamma$ and $\omega=0$, as a function of the kinetic-energy cutoff $E_k$ for the plane waves of the DFT ground-state calculation, and in Table~\ref{tab:mono_Sex_vs_ecutscf} the convergence of $\Sigma_x (\mathbf{k})$ for the three electronic states at $\mathbf{k}=\Gamma$ closest to the Fermi energy. The convergence with respect to the kinetic-energy cutoff is quite fast, allowing us to choose a value of $40\;$Ry to achieve an accuracy $<100\;$meV.

\begin{table}[h!]
	\begin{center}
		\caption{The screened Coulomb interaction $W_{i} = W_{\mathbf{G}_i,\mathbf{G}_i}(\mathbf{q},\omega)$ with $\mathbf{G}_i = (i,0,0) 2\pi/a$, evaluated at $\mathbf{q}=\Gamma$ and $\omega=0$, as a function of the kinetic energy cutoff $E_k$ for the plane waves of the DFT ground state calculation. For these calculations, we used a $24 \times 24 \times 1$ $\mathbf{k}$-grid, an energy cutoff in the dielectric matrix of $10\;$Ry and an energy cutoff of $20\;$Ry in the exchange part of the self-energy.}
		\label{tab:mono_W_vs_scfecut}
		\begin{tabular}{c|c|c|c} 
    $E_k$ (Ry) & $W_{2}$ & $W_{4}$ & $W_{6}$  \\ \hline
			20 & -0.6721 & -0.2595 & -0.2423  \\ \hline
			30 & -0.6741 & -0.2600 & -0.2433  \\ \hline
			40 & -0.6737 & -0.2598 & -0.2430  \\ \hline
			50 & -0.6735 & -0.2598 & -0.2429  \\ \hline
			60 & -0.6734 & -0.2597 & -0.2428  \\ \hline
		\end{tabular}
	\end{center}
\end{table}

\begin{table}[h!]
	\begin{center}
		\caption{$\Sigma_x (\mathbf{k})$ of the three electronic states at $\mathbf{k}=\Gamma$ closest to the Fermi energy (at DFT level) as a function of the kinetic-energy cutoff $E_k$ for the plane-waves of the DFT ground state calculation. For these calculations, we used a $12 \times 12 \times 1$ $\mathbf{k}$ grid and an energy cutoff in the dielectric matrix of $10 \;$Ry.}
		\label{tab:mono_Sex_vs_ecutscf}
		\begin{tabular}{c|c|c|c} 
	   $E_k$ (Ry) & band 1 (eV) & band 2 (eV) & band 3 (eV) \\ \hline
			20    & -16.03      & -9.82       & -10.29 \\ \hline
			30 	  & -16.04      & -9.87       & -10.34 \\ \hline
			40 	  & -16.03      & -9.84       & -10.30 \\ \hline
			50 	  & -16.02      & -9.82       & -10.29 \\ \hline
			60 	  & -16.02      & -9.81       & -10.28 \\ \hline
		\end{tabular}
	\end{center}
\end{table}

\subsection{Brillouin zone sampling}
\label{app:bz_mono}
\begin{figure}
  	\includegraphics[width=1.0\columnwidth]{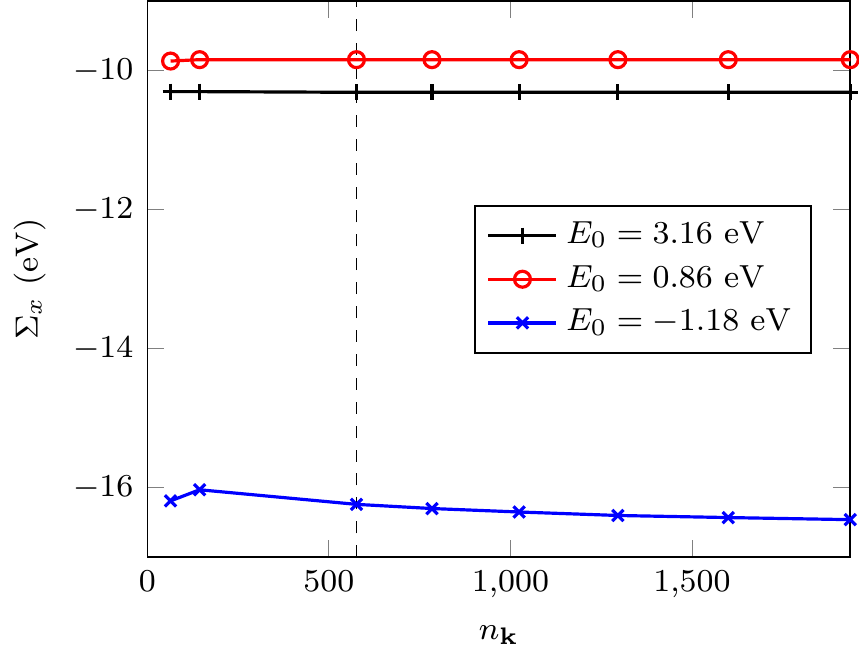}
  	\caption{$\Sigma_x (\mathbf{k})$ of the three electronic states at $\mathbf{k}=\Gamma$ closest to the Fermi energy (at DFT level) as a function of the number of BZ points $n_{\mathbf{k}}$. The vertical dashed line indicates the value for the $24 \times 24 \times 1$ BZ grid used in the final calculations.}
  	\label{fig:mono_Sx_vs_nk}
\end{figure}

In Table~\ref{tab:mono_W_vs_nk} we report the convergence of diagonal elements of the screened Coulomb interaction $W_{i} = W_{\mathbf{G}_i,\mathbf{G}_i}(\mathbf{q},\omega)$ with $\mathbf{G}_i = (i,0,0) 2\pi/a$, evaluated at $\mathbf{q}=\Gamma$ and $\omega=0$, as a function of the number of BZ points $n_\mathbf{k}$ used to sample the Brillouin zone, which shows a very favorable behavior.
\begin{table}[h!]
	\begin{center}
		\caption{Diagonal elements of the screened Coulomb interaction $W_{i} = W_{\mathbf{G}_i,\mathbf{G}_i}(\mathbf{q},\omega)$ with $\mathbf{G}_i = (i,0,0) 2\pi/a$, evaluated at $\mathbf{q}=\Gamma$ and $\omega=0$, as a function of the number of BZ-points $n_\mathbf{k}$ used to sample the Brillouin zone. For these calculations, we used an energy cutoff in the dielectric matrix of $10\;$Ry and an energy cutoff of $20\;$Ry in the exchange part of the self-energy.}
		\label{tab:mono_W_vs_nk}
		\begin{tabular}{c|c|c|c} 
$n_\mathbf{k}$ (Ry) & $W_{2}$ & $W_{4}$ & $W_{6}$  \\ \hline
		      $8^2$ & -0.6788 & -0.2600 & -0.2431  \\ \hline
			 $12^2$ & -0.6711 & -0.2596 & -0.2429  \\ \hline
			 $16^2$ & -0.6723 & -0.2597 & -0.2430  \\ \hline
			 $20^2$ & -0.6736 & -0.2598 & -0.2430  \\ \hline
			 $24^2$ & -0.6737 & -0.2598 & -0.2430  \\ \hline
			 $28^2$ & -0.6737 & -0.2599 & -0.2430  \\ \hline
			 $32^2$ & -0.6738 & -0.2599 & -0.2430  \\ \hline
			 $36^2$ & -0.6738 & -0.2599 & -0.2430  \\ \hline
		\end{tabular}
	\end{center}
\end{table}
Converging the exchange part of the self-energy with respect to the number of BZ points is less smooth than for the correlation part, as already noticed for the bulk. In contrast to the bulk, however, we find that the slowest convergence is not observed for the state crossing the Fermi level, but for the fully occupied band below, as can be seen in Fig.~\ref{fig:mono_Sx_vs_nk}. The difference in $\Sigma_x (\mathbf{k}=\Gamma)$ for this electronic state between the chosen $24 \times 24 \times 1$ BZ grid and a $44 \times 44 \times 1$ grid, considered converged, is around $200\;$meV. As this electronic band is fully occupied and will only be pushed further down in energy, the Fermi surface, which is our main point of focus, will not change and we accept a larger inaccuracy for this state in view of reducing computational time.

\subsection{Energy cutoff $E_c$ for the correlation part of the self-energy $\Sigma_c$}
\label{app:diel_mono}
In Table~\ref{tab:mono_W_vs_ecutsco} we report the convergence of diagonal elements of the screened Coulomb interaction \mbox{$W_{i} = W_{\mathbf{G}_i,\mathbf{G}_i}(\mathbf{q},\omega)$} with $\mathbf{G}_i = (i,0,0) 2\pi/a$ evaluated at $\mathbf{q}=\Gamma$ and $\omega=0$ as a function of the energy cutoff $E_c$ for the dielectric matrix. 

\begin{table}[h!]
	\begin{center}
		\caption{Diagonal elements of the screened Coulomb interaction $W_{i} = W_{\mathbf{G}_i,\mathbf{G}_i}(\mathbf{q},\omega)$ with $\mathbf{G}_i = (i,0,0) 2\pi/a$, evaluated at $\mathbf{q}=\Gamma$ and $\omega=0$, as a function of the energy cutoff $E_c$ for the dielectric matrix. For these calculations, we used a $24 \times 24 \times 1$ $\mathbf{k}$ grid and a plane-wave cutoff of $40\;$Ry.}
		\label{tab:mono_W_vs_ecutsco}
		\begin{tabular}{c|c|c|c}
			$E_c$ (Ry) & $W_{2}$     & $W_{4}$     & $W_{6}$  \\ \hline
			 6         & -0.6575     &  -0.2487    &  -0.2043   \\ \hline
			 8         & -0.6571     &  -0.2485    &  -0.2042   \\ \hline
			10         & -0.6569     &  -0.2484    &  -0.2041   \\ \hline
			12         & -0.6568     &  -0.2484    &  -0.2041   \\ \hline
			14         & -0.6568     &  -0.2484    &  -0.2041   \\ \hline
		\end{tabular}
	\end{center}
\end{table}

In Table~\ref{tab:mono_bands_vs_ecutsco} we show the convergence behavior of the $GW$ quasiparticle energies at $\Gamma$ closest to the DFT Fermi energy as a the energy cutoff $E_c$ for the dielectric matrix. Convergence is very fast, allowing us to choose a value of $10\;$Ry for the energy cutoff $E_c$.

\begin{table}[h!]
    \begin{center}
        \caption{Quasiparticle eigenvalues of the three electronic states at $\Gamma$ closest to the Fermi energy (at DFT level) as a function of the energy cutoff $E_c$ for the dielectric matrix. For these calculations, we used an $12 \times 12 \times 1$ $\mathbf{k}$ grid, a plane-wave cutoff of $40\;$Ry, and an energy cutoff for the exchange part of 20~Ry.}
        \label{tab:mono_bands_vs_ecutsco}
        \begin{tabular}{c|c|c|c}
            $E_c$ (Ry) & band 1 (eV) & band 2 (eV) & band 3 (eV) \\ \hline
            6          & -2.77       &  0.89       &  3.29       \\ \hline
            8          & -2.78       &  0.65       &  2.93       \\ \hline
            10         & -2.81       &  0.44       &  2.71       \\ \hline
            12         & -2.87       &  0.30       &  2.57       \\ \hline
            14         & -2.92       &  0.19       &  2.46       \\ \hline
            16         & -2.94       &  0.16       &  2.42       \\ \hline
        \end{tabular}
    \end{center}
\end{table}

\subsection{Energy cutoff $E_x$ for the exchange part of the self-energy $\Sigma_x$}
\label{app:ex_mono}
In Table~\ref{tab:mono_Sex_vs_ecutsex} we report the convergence of $\Sigma_x (\mathbf{k})$ of the three electronic states at $\Gamma$ closest to the Fermi energy (at DFT level) as a function of the energy cutoff $E_x$. We find again that not the electronic state crossing the Fermi level, but the one below, shows the slowest convergence. Still, by choosing $E_x = 25\;$Ry, we can achieve an accuracy of $<50\;$meV.

\begin{table}[h!]
	\begin{center}
		\caption{$\Sigma_x (\mathbf{k})$ of the three electronic states at $\mathbf{k}=\Gamma$ closest to the Fermi energy (at DFT level) as a function of the energy cutoff $E_x$. For these calculations, we used a $12 \times 12 \times 1$ $\mathbf{k}$-grid, an energy cutoff in the dielectric matrix of $10 \;$Ry and a plane wave cutoff of $40 \;$Ry.}
		\label{tab:mono_Sex_vs_ecutsex}
		\begin{tabular}{c|c|c|c} 
			$E_x$ & band 1 (eV) & band 2 (eV) & band 3 (eV) \\ \hline
			15    & -15.80  & -9.47  &  -9.97 \\ \hline
			20 	  & -16.03  & -9.84  & -10.30 \\ \hline
			25 	  & -16.11  & -9.93  & -10.40 \\ \hline
			30 	  & -16.14  & -9.96  & -10.43 \\ \hline
			35 	  & -16.14  & -9.97  & -10.44 \\ \hline
		\end{tabular}
	\end{center}
\end{table}



\bibliographystyle{apsrev4-1}
%

\end{document}